\documentclass[aps,preprint,showpacs,groupedaddress,floatfix]{revtex4}

\usepackage{graphicx}
\usepackage{dcolumn}
\usepackage{bm}

\begin{document}
%\begin{CJK*}{GBK}{song}

\title{The Gamow-Teller response within Skyrme random-phase approximation plus particle-vibration coupling}
\author{Y. F. Niu$^{1,2}$}
\author{G. Col\`{o} $^{2}$}
\email{gianluca.colo@mi.infn.it}
\author{M. Brenna$^2$}
\author{P.F. Bortignon $^2$}
\author{J. Meng$^{1,3,4}$}

 \affiliation{$^1$ State Key Laboratory of
Nuclear Physics and Technology, School of Physics, Peking
University, Beijing 100871, China}
 \affiliation{$^2$ Dipartmento di Fisica, Unversit\`{a} degli Studi
di Milano,
 and INFN, Sezione di Milano, I-20133 Milano, Italy}
 \affiliation{$^3$ School of Physics and Nuclear Energy Engineering,
Beihang University, Beijing 100191, China}
 \affiliation{$^4$ Department of Physics, University of Stellenbosch, Stellenbosch 7602, South Africa}

\date{\today}
\begin{abstract}

Although many random-phase approximation~(RPA) calculations of the
Gamow-Teller (GT) response exist, this is not the case for
calculations going beyond the mean-field approximation. We apply a
consistent model that includes the coupling of the GT resonance to
low-lying vibrations, to nuclei of the $fp$ shell. Among other
motivations, our goal is to see if the particle-vibration coupling
can redistribute the low-lying GT$^+$ strength that is relevant for
electron-capture processes in core-collapse supernova. We conclude
that the lowering and fragmentation of that strength are consistent
with the experimental findings and validate our model. However, the
particle-vibration coupling cannot account for the quenching of the
total value of the low-lying strength.

\end{abstract}
 \maketitle
\date{today}
%%%%%%%%%%%%%%%%%%%%%%%%%%%%%%%%%%%%%%%

\section{Introduction}

The Gamow-Teller (GT) resonances are among the clear
manifestations of nuclear collective motion. In (p,n)
or ($^3$He, t) reactions studied at small angles or
zero angular momentum transfer, when the target nucleus
is perturbed by an external field proportional to
\begin{equation}
\hat O_{\rm GT ^-} = \sum_{i=1}^A \vec \sigma(i) \cdot \tau_-(i),
\end{equation}
a broad peak systematically appears in all the nuclei that have been
studied in the last three decades~\cite{Osterfeld1992}. The position
and the magnitude of such a peak can be explained only by assuming a
coherent effect among particle-hole (p-h) excitations in the
spin-isospin channel.

Accordingly, much effort has been spent in trying to understand if
the systematic study of the GT response can well constrain the
nuclear interaction in the spin-isospin
channel~\cite{Bender2002,Paar2004,Paar2007,Fracasso2007,Liang2008,Bai2009},
which will be discussed in this work as well. From the nuclear
structure point of view, another interesting phenomenon is the
so-called Gamow-Teller quenching. The difference between the total
strength associated, respectively, with the $\hat O_{\rm GT^-}$ and
$\hat O_{\rm GT^+}$ operators [the latter being of course analogous
to the former one but proportional to $\tau_+(i)$], has the
model-independent value given by $3(N-Z)$, which is the well-known
Ikeda sum rule~\cite{Ikeda1963} and will be denoted below by
$m_0({\rm GT^-}) - m_0({\rm GT^+})$. Only about 60\%--70\% of this
value has been found around the main GT peak~\cite{Gaarde1983}. In
$^{90}$Zr, it has been possible to assess the relative importance of
different possible quenching mechanisms~\cite{Wakasa1997}. In
principle, the quenching may be due either to coupling with
configurations that are more complicated than the simple p-h
ones~\cite{Bertsch1982,Drozdz1986}, or to coupling with internal
degrees of freedom of the nucleon as with the $\Delta$ particle.
Since $93\%$ of the Ikeda sum rule is found below $50$
MeV~\cite{Wakasa1997}, not much room is left for the latter
mechanism.

At the same time the Gamow-Teller and the other spin-isospin
charge-exchange nuclear transitions do play a relevant role for
particle physics~\cite{Towner2010,Liang2009} and
astrophysics~\cite{Langanke2003}. We will not discuss much their
relevance for single and double $\beta$ decay, but we will focus on
the connection between GT transitions and the evolution of massive
stars at the end of the last hydrostatic burning phase. In fact, if
the mass of the iron core exceeds the so-called Chandrasekar mass,
then the pressure of the degenerate electron gas is not sufficient
to make the system stable against gravitational collapse. In this
scenario, one of the main processes that governs the subsequent
evolution of the star until an eventual supernova explosion is the
electron capture both by free protons and by core nuclei in the iron
region~\cite{Bethe1979,Bethe1990,Langanke2003,Janka2007}.

Extensive tables of electron capture rates, based on realistic
estimates of electron capture cross sections, would be needed for
supernova simulations. At the electron energies of interest,
typically less than 30 MeV, electron capture is dominated by the
GT$^+$ contribution in nuclei around the mass region of
Fe~\cite{Dean1998,Langanke2001,Paar2009,Niu2011}. When an accurate
experimental determination of the corresponding strength is missing,
one is obliged to resort to theoretical frameworks such as the
shell-model (SM) and the mean-field (MF) or energy-density
functional (EDF) based schemes. EDF-based calculations of the
electron capture process are worthwhile for several reasons. They
are less demanding than SM calculations and are not limited to $sd$
or $fp$ shells. A cross-comparison with the SM can help in reducing
the uncertainties associated with the implementation of the
EDF-based scheme, and, in a complementary way, the same comparison
is certainly instrumental to validate different EDF parameter sets,
in the path toward a universal EDF.

In Ref.~\cite{Paar2009}, the electron capture cross sections on
several nuclei have been calculated by using a self-consistent
implementation of the Skyrme Hartree-Fock plus random-phase
approximation (HF plus RPA) model. This model, well known and widely
used for many years, has been for the first time extended to finite
temperature. The results have been compared with SM results from
Refs.~\cite{Dean1998,Langanke2001}.

The two models predict cross sections that are not very different at
energies around $10-20$ MeV. At high energies, the SM predictions
lie below the HF-RPA ones, whereas at low energies (below about $10$
MeV) the situation is reversed. In fact, the main drawback of HF-RPA
seems to be the fact that it predicts a too-high threshold energy
for the GT$^+$. Another feature of the finite temperature HF-RPA
calculations is a certain spread of the results associated with
different parameter sets.

The present work is aimed to increase understanding and try to
overcome these open problems. In the simpler zero-temperature case,
where one can compare with experimental measurements of the GT$^+$
strength, we first study the underlying reasons for the sensitivity
of this quantity to the choice of the Skyrme parameter set. Then, we
analyze the problem of the threshold energy and the fragmentation of
the GT strength. The Skyrme functionals are well known to be unable
to describe the single-particle states around the Fermi energy: they
tend to predict a too-large single-particle gap, and cannot by
definition reproduce the fragmentation of the single-particle
strength. As the particle-vibration coupling (PVC) approach is one
possible improvement in this respect, we will study to which extent
a model based on RPA plus particle-vibration coupling can better
reproduce the GT strength in the nuclei of interest.

The plan of our paper is therefore the following. We provide first a
description of the basic formalism in Sec. \ref{formalism} as well
as of the numerical input in Sec. \ref{num}. The results concerning
the sensitivity to the choice of the Skyrme set are discussed in
Sec. \ref{Skyrme}, and the results of RPA plus particle-vibration
coupling are discussed in Sec. \ref{PVC}. We mainly concentrate on
the nucleus $^{60}$Ni, as a typical system in the mass region of
interest. Meanwhile, we also show some results for other nuclei such
as $^{56}$Ni, which has been object of a recent experimental study
\cite{Sasano2011}. We present conclusions in Sec. \ref{conclu}.

\section{Sketch of the Theoretical Framework}\label{formalism}
In the present work, we use the same basic formalism already
employed in Ref. \cite{Colo1994}. The main difference is that the
present PVC calculation is more consistent as we discuss below. The
first step is a HF plus RPA calculation of the Gamow-Teller
strength. This is done in the same way as in Ref.
\cite{Fracasso2007}. The second step consists of implementing the
coupling with vibrations.

We start from the solution of the HF equations for a given nucleus
$(A,Z)$ using a Skyrme two-body interaction. Then, we set up a
discrete basis of both proton-neutron and neutron-proton p-h
configurations. The continuum is discretized by putting the system
in a box and requiring vanishing boundary conditions for the wave
functions at the surface of this box. We write the RPA matrix
equations and solve them in the model space of the p-h discrete
configurations, which is called $Q_1$. These equations are well
known from textbooks~\cite{RingBook} and we shall not discuss them
further here. The RPA provides, as a rule, an accurate description
of the centroid of giant resonances and the fraction of the energy
weighted sum rule exhausted by the mode.

Giant resonances have a quite large damping width. In order to
explain it, one needs to go beyond the RPA. As discussed, e.g., in
Ref.~\cite{Colo1994}, the width of giant resonances is mainly made
up of an escape width ($\Gamma^\uparrow$) and a spreading width
($\Gamma^ \downarrow$). These are due to different mechanisms,
namely to the decay through particle emission and to the coupling
with more complicated states of the nuclear spectrum, respectively.
The RPA can reproduce the escape width if proper coupling to the
continuum is implemented, but cannot account for the spreading
width.

Here we will follow the formalism of Ref.~\cite{Colo1994}, and only
focus on the spreading width which is the leading damping mechanism.
The processes in which the energy and angular momentum associated
with the vibrational nuclear motion are distributed among more
complex internal degrees of freedom do contribute to the spreading
width. In order to account for these effects we need a subspace,
which we shall call $Q_2$ and which is built with a set of ``doorway
states.'' We denote these doorway states by $|N\rangle$, and make a
physical choice of them in terms of states made up with a p-h
excitation coupled to a collective vibration. Note that we use the
same Skyrme interaction to calculate both the GT states and all the
ingredients to build up the space $Q_2$, namely we employ Skyrme-HF
single-particle states and vibrations that are calculated
consistently in the RPA with the same Skyrme set. In this respect,
our scheme is parameter free.

We use the projection formalism to restrict our
effective Hamiltonian to the subspace $Q_1$ and make
the calculations feasible. After truncation of higher-order
couplings, the effective Hamiltonian reads
 \begin{equation}\label{Heff}
   {\cal H} (\omega) = Q_1 H Q_1   + W^\downarrow
  (\omega) = Q_1HQ_1   + Q_1 H Q_2 \frac{1}{\omega-Q_2HQ_2+i\epsilon}
  Q_2HQ_1,
 \end{equation}
where $\omega$ is the excitation energy. This energy-dependent,
complex Hamiltonian has complex eigenvalues whose imaginary parts
originate from the coupling to the more complicated configurations.
We shall denote these complex states as $\vert\nu\rangle$ in what
follows.

In practice, we first diagonalize the RPA Hamiltonian $Q_1 H Q_1$
and obtain the complete basis made up with the RPA states. The
creation operators ${\cal O}_\nu^\dagger$ of the states
$|\nu\rangle$ can be expressed as a linear combination of the RPA
creation operators, namely
 \begin{equation}
  {\cal O}^\dagger_\nu = \sum_{\omega_n>0} F_n^{(\nu)} O_n^\dagger -
  \bar{F}_n ^{(\nu)} \bar{O}_n^\dagger,
 \end{equation}
 where $O_n^\dagger$ and $\bar{O}_n^\dagger$ are creation operators
 of the RPA states $|n\rangle$ lying at positive energy $\omega_n$,
and of the states $|\bar{n}\rangle$ at the corresponding negative energy
 $-\omega_n$, respectively. Then the eigenvalue equation for the effective Hamiltonian (\ref{Heff}), that is,
 \begin{equation}
  [{\cal H}, {\cal O}^\dagger_\nu ] = (\Omega_\nu -i
  \frac{\Gamma_\nu}{2}) {\cal O}^\dagger_\nu,
 \end{equation}
can be cast in matrix form on the RPA basis as
 \begin{equation}
 \label{RPAmatrix}
    \left( \begin{array}{cc} {\cal D} + {\cal A}_1(\omega) & {\cal
    A}_2(\omega) \\ -{\cal A}_3(\omega) & -{\cal D} - {\cal
    A}_4(\omega) \end{array} \right) \left( \begin{array}{c}
    F^{(\nu)} \\ \bar{F}^{(\nu)} \end{array} \right) = (\Omega_\nu - i
    \frac{\Gamma_\nu}{2}) \left( \begin{array}{c}
    F^{(\nu)} \\ \bar{F}^{(\nu)} \end{array} \right).
   \end{equation}
In this latter equation ${\cal D}$ is a diagonal matrix with the
positive RPA eigenvalues, and the ${\cal A}_i$ matrices contain the
elements associated with the second term of Eq.~(\ref{Heff}) denoted
by $W^\downarrow$. The matrix $\left( \begin{array}{cc} {\cal D} +
 {\cal A}_1(\omega) & {\cal
    A}_2(\omega) \\  {\cal A}_3(\omega) &  {\cal D} + {\cal
    A}_4(\omega) \end{array} \right) $
is complex and symmetric, as it can be seen from their explicit form
provided below. The orthogonality and normalization relations of the
eigenvectors are
 \begin{equation}
 \sum_{n} F^{(\nu)}_{n}F^{(\nu')}_{n} - \bar{F_{n}}^{(\nu)} \bar{F_{n}}^{(\nu')} = \delta_{\nu\nu'}.
 \end{equation}

Our goal is to extract from the diagonalization of the effective
Hamiltonian (\ref{Heff}) a physical observable. In particular, we
are naturally interested in the response function associated with an
external operator such as the GT operator defined in the
Introduction. This response function can be written as
 \begin{equation}
 \label{response}
  R(\omega) = \langle 0 | \hat O_{\rm GT}^\dagger
\frac{1}{\omega - {\cal H}(\omega) +
  i\eta} \hat O_{\rm GT} | 0 \rangle.
 \end{equation}
We use a notation that includes both the GT$^-$ and GT$^+$ cases
(one should note that $\hat O_{\rm GT^-}^\dagger$ is equal to $\hat
O_{\rm GT^+}$, and vice versa). The strength function is related to
Eq. (\ref{response}) by the well-known relation
 \begin{equation}
  S(\omega) = -\frac{1}{\pi} {\rm Im} R(\omega)
  = -\frac{1}{\pi} {\rm Im} \sum_\nu \langle 0 | \hat O_{\rm GT} | \nu \rangle
  ^2 \frac{1}{\omega - \Omega_\nu + i\frac{\Gamma_\nu}{2}},
 \end{equation}
where the squared matrix element of the transition operator
appears, and not its
squared modulus, due to the properties of the eigenvectors
$|\nu\rangle$ which form a biorthogonal basis.

\begin{figure}
\centerline{
\includegraphics[scale=0.65,angle=0]{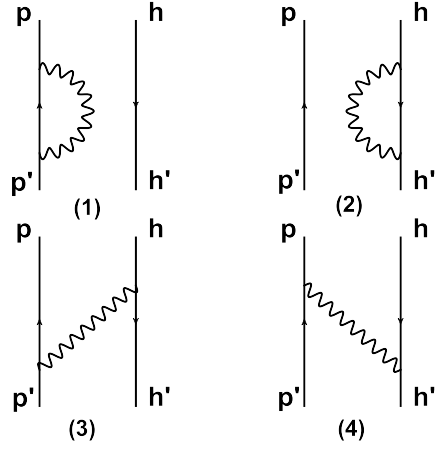}
} \caption{Diagrammatic representation of the four terms whose sum
gives the matrix element $W^\downarrow _{ph,p'h'}$. The analytic
expressions are shown in Eq. (\ref{Wdown1}).} \label{fig1}
\end{figure}

 We now provide some more details on the
calculation of the matrix elements of $W^\downarrow$. On the basis
of the p-h configurations in $Q_1$ space on which the RPA has been
solved, the matrix element of $W^\downarrow (\omega)$ will be
denoted as $W^{\downarrow}_{ph,p'h'} (\omega)$. If these matrix
elements are known, then the matrix elements ${\cal A}_i$ on the RPA
basis are obtained through a straightforward basis transformation,
 \begin{eqnarray}
  ({\cal A}_1) _{mn} &=& \sum_{ph,p'h'} W^\downarrow _{ph,p'h'} (\omega) X_{ph}^{(m)} X^{(n)}_{p'h'}
  + W^{\downarrow*} _{ph,p'h'} (-\omega) Y_{ph}^{(m)} Y^{(n)}_{p'h'}, \\
  ({\cal A}_2) _{mn} &=& \sum_{ph,p'h'} W^\downarrow _{ph,p'h'} (\omega) X_{ph}^{(m)} Y^{(n)}_{p'h'}
  + W^{\downarrow*} _{ph,p'h'} (-\omega) Y_{ph}^{(m)} X^{(n)}_{p'h'}, \\
  ({\cal A}_3) _{mn} &=& \sum_{ph,p'h'} W^\downarrow _{ph,p'h'} (\omega) Y_{ph}^{(m)} X^{(n)}_{p'h'}
  + W^{\downarrow*} _{ph,p'h'} (-\omega) X_{ph}^{(m)} Y^{(n)}_{p'h'}, \\
  ({\cal A}_4) _{mn} &=& \sum_{ph,p'h'} W^\downarrow _{ph,p'h'} (\omega) Y_{ph}^{(m)} Y^{(n)}_{p'h'}
  + W^{\downarrow*} _{ph,p'h'} (-\omega) X_{ph}^{(m)} X^{(n)}_{p'h'}.
  \end{eqnarray}
$W^{\downarrow}$ is clearly given by
  \begin{equation}
  W^{\downarrow}_{ph,p'h'} (\omega) = \sum_N \frac{\langle ph |V|N\rangle \langle N | V | p'h' \rangle }{\omega
  -\omega_N}.
 \end{equation}
The matrix elements at the numerator can be evaluated using Wick's
theorem. $W^{\downarrow}_{ph,p'h'} (\omega)$ turns out to be the sum
of the four terms whose diagrammatic representation is shown in
Fig.~\ref{fig1} and whose analytic expression is
 \begin{eqnarray}
 \label{Wdown1}
 W^{\downarrow }(1) &=&\delta_{hh'}\delta_{j_p j_{p'}} \sum_{p'',nL}
  \frac{1} {\omega-(\omega_n+\epsilon_{p''}-\epsilon_h)+i\eta}
  \frac{  \langle p || V || p'',nL
  \rangle \langle p' || V || p'',nL
  \rangle }{\hat{j}_p^2} ,\nonumber\\
  W^{\downarrow }(2) &=& \delta_{ p {p'}}   \delta_{j_h j_{h'}} \sum_{h'',nL}
  \frac{1} {\omega-(\omega_n-\epsilon_{h''}+\epsilon_p)+i\eta}\frac{  \langle h || V || h'',nL
  \rangle \langle h' || V || h'',nL
  \rangle }{\hat{j}_h ^2} ,\nonumber\\
  W^{\downarrow }(3) &=& \sum_{nL}
  \frac{(-)^{ j_{p}-j_{h'}+J+L}}{\omega-(\omega_n+\epsilon_{p}-\epsilon_{h'})+i\eta}
  \left\{ \begin{array}{ccc} j_p & j_h & J \\ j_{h'} &
  j_{p'} & L \end{array} \right\}
   \langle p' || V || p, nL  \rangle \langle h' || V || h,nL
   \rangle, \nonumber\\
   W^{\downarrow }(4)&=& \sum_{nL}
  \frac{(-)^{ j_{p'}-j_{h}+J+L}}{\omega-(\omega_n+\epsilon_{p'}-\epsilon_{h})+i\eta}
  \left\{ \begin{array}{ccc} j_p & j_h & J \\ j_{h'} &
  j_{p'} & L \end{array} \right\}
   \langle p || V || p',nL \rangle  \langle h || V || h' , nL
  \rangle.\nonumber\\
 \end{eqnarray}
In the above formulas, $p$ and $h$ label always particle and hole
states, respectively. The corresponding angular momentum and
single-particle energy are given by $j_i$ and $\epsilon_i$.
${\hat{j}_i^2}$ is a shorthand notation for $2j_i+1$. The phonon
states are labeled by their angular momentum $L$ (only natural
parity states are included in our calculations) and by an additional
index $n$. The (small) parameter $\eta$ is introduced to mimic
couplings beyond the doorway-states approximation: this parameter is
set at the value of $500$ keV.

%---------------------------------------------------------------------------------------------------------
%----------------------------------------------------------------------------------------------------------

Up to this point, the scheme is exactly the same as in
Ref.~\cite{Colo1994}. In Ref.~\cite{Colo1994}, the interaction at
the particle-vibration coupling vertex [$V$ in Eq. (\ref{Wdown1})]
was approximated by retaining only the $V_0$ part in the following
momentum-independent terms of the Skyrme p-h force ($t_0$ and $t_3$
terms),
 \begin{eqnarray}
  V_{qq} &=& V_0^{qq} \delta(\boldsymbol{r}_1 - \boldsymbol{r}_2) + V_\sigma^{qq} \delta(\boldsymbol{r}_1 -
  \boldsymbol{r}_2) \boldsymbol{\sigma}_1 \cdot
  \boldsymbol{\sigma}_2,\nonumber\\
  V_{q\tilde{q}}& =& V_0^{q\tilde{q}} \delta(\boldsymbol{r}_1 - \boldsymbol{r}_2) + V_\sigma^{q\tilde{q}} \delta(\boldsymbol{r}_1 -
  \boldsymbol{r}_2) \boldsymbol{\sigma}_1 \cdot
  \boldsymbol{\sigma}_2.
  \end{eqnarray}
  The functions $V_0$ and $V_\sigma$ depend only on the radial coordinate $r$, and
  their detailed expressions are
 \begin{eqnarray}
  V_0^{qq}(r) &=& \frac{1}{2} t_0(1-x_0) + \frac{1}{16} t_3 (\alpha
  +2) (\alpha +1 ) \rho^\alpha(r) - \frac{1}{12}t_3(x_3 +
  \frac{1}{2})\rho^\alpha(r) \nonumber\\
  && + \frac{1}{48}t_3\alpha(1-\alpha)(1+2x_3) \rho^{\alpha-2}(r)
  \rho_-^2(r) - \frac{1}{12} t_3 (2x_3+1) \alpha \rho^{\alpha-1}(r)
  \rho_-(r),\nonumber\\
  V_0^{q\tilde{q}}(r) &=& \frac{1}{2} t_0(2+x_0) + \frac{1}{16} t_3 (\alpha
  +2) (\alpha +1 ) \rho^\alpha(r) + \frac{1}{12}t_3(x_3 +
  \frac{1}{2})\rho^\alpha(r) \nonumber\\
  && + \frac{1}{48}t_3\alpha(1-\alpha)(1+2x_3) \rho^{\alpha-2}(r)
  \rho_-^2(r),\nonumber\\
  V_\sigma^{qq}(r) &=& \frac{1}{2}t_0(x_0-1) - \frac{1}{12}
  t_3(1-x_3)\rho^\alpha(r),\nonumber\\
  V_\sigma^{q\tilde{q}}(r) &=& \frac{1}{2}t_0x_0 + \frac{1}{12}
  t_3x_3\rho^\alpha(r).
  \end{eqnarray}
In these formulas, $q$ and $\tilde{q}$ label different charge states
of two p-h pairs, where $q\neq \tilde{q}$. $\rho_-=\rho_n-\rho_p$
stands for the density difference of neutron and proton, while
$\rho=\rho_n+\rho_p$ is the total nucleon density. The other symbols
are the standard Skyrme parameters. In the present work, we include
the whole central Skyrme p-h force, i.e., not only all the $t_0$ and
$t_3$ terms, but also the $t_1$ and $t_2$ terms with the following
form:
\begin{eqnarray}
 V &=& \frac{1}{2} t_1(1+x_1 P_\sigma) [\boldsymbol{P}^{'2} \delta(\boldsymbol{r}_1-\boldsymbol{r}_2)
 + \delta(\boldsymbol{r}_1-\boldsymbol{r}_2)\boldsymbol{P}^{2}  ] (1-P_\sigma
 P_\tau) \nonumber\\
 && + t_2(1+x_2 P_\sigma) \boldsymbol{P}' \cdot
 \delta(\boldsymbol{r}_1-\boldsymbol{r}_2) \boldsymbol{P}(1+P_\sigma P_\tau),
\end{eqnarray}
where $\boldsymbol{P} = \frac{1}{2i} (\nabla_1 - \nabla_2)$, and
$\boldsymbol{P}'$ is the Hermitian conjugate of $\boldsymbol{P}$
(acting on the left). $P_\sigma = \frac{1}{2} (1+
\boldsymbol{\sigma}_1\cdot \boldsymbol{\sigma}_2)$ and $P_\tau=
\frac{1}{2} (1+ \boldsymbol{\tau}_1\cdot \boldsymbol{\tau}_2)$ are
the spin-exchange and isospin-exchange operators.
 This is consistent with the findings of
Ref.~\cite{Colo2010}, where it has been shown that the
momentum-dependent terms have a non-negligible effect. In
Ref.~\cite{Colo2010} the expression of the reduced matrix elements
of $V$ is derived. We report the result here, that is,
 \begin{equation}
  \langle i || V || j, nL
 \rangle =  \sqrt{2L+1} \sum_{ph}
   \left[X_{ph}^{nL} V_L(ihjp) + (-)^{L+j_h-j_p} Y_{ph}^{nL} V_L(ipjh)\right],
  \end{equation}
  where the RPA amplitudes appear and $V_L$ is the p-h coupled matrix element,
  \begin{equation}
  V_L(ihjp)=\sum_{\text{all }m} (-)^{j_j-m_j+j_h-m_h} \langle j_i
  m_i j_j -m_j | LM \rangle  \langle j_p
  m_p j_h -m_h | LM \rangle \langle j_i m_i , j_h m_h | V | j_j
  m_j,j_pm_p\rangle.
 \end{equation}

We briefly recall, for the reader's convenience, the relationship
between the present approach in which the full central part of the
Skyrme interaction is considered in the PVC vertex, and the
approximate one that was used in the past. If, as in
Ref.~\cite{Colo1994}, only the $V_0(r) \delta(\boldsymbol{r}_1 -
\boldsymbol{r}_2)$ part in $t_0$ and $t_3$ terms is included in the
interaction $V$, the p-h coupled matrix elements have the simple
form
 \begin{eqnarray}
  V_L(ihjp) &=& \frac{i^{-l_i-l_h+l_j+l_p}}{2L+1} \langle i || Y_L
  || j\rangle \langle p || Y_L || h \rangle \int \frac{dr}{r^2}
  V_{0}^{qq'}(r) u_i(r) u_j(r) u_p(r) u_h(r), \nonumber \\
  V_L(ipjh)& =& (-)^{L+j_p-j_h} V_L(ihjp),
 \end{eqnarray}
 where the radial part of the s.p. (single particle) wave functions, written as
$\phi_{nljm}=\frac{u_{nlj}(r)}{r} [Y_l \otimes
 \chi_{1/2}]_{jm}$, has been introduced. Thus,
 \begin{eqnarray}
 \label{Vph}
 \langle i || V || j,nL \rangle &=& \sqrt{2L+1} \sum_{ph}( X^{nL}_{ph}+Y^{nL}_{ph})
 V_L(ihjp) \nonumber \\
 &=& \langle i || Y_L
  || j\rangle \sum_{q'} \int dr V^{qq'}_0(r) u_i(r) u_j(r) \delta
  \rho_{nL}^{(q')},
  \end{eqnarray}
where $q$ labels the charge of the states $i$ and $j$, and
the neutron or proton radial transition density of the state $|nL\rangle$
has been introduced. This is
 \begin{equation}
  \delta\rho_{nL}^{(q)} (r) = \frac{1}{\sqrt{2L+1}} \sum_{ph\in q}( X^{nL}_{ph}+Y^{nL}_{ph})\langle p || Y_L || h \rangle
  \frac{u_p(r)}{r} \frac{u_h(r)}{r}.
  \end{equation}
According to the further approximation used in Ref.~\cite{Colo1994},
i.e., namely that a collective state should have mainly isoscalar
(or isovector) character so that its isovector (isoscalar)
transition density can be neglected, Eq. (\ref{Vph}) takes the even
simpler form
 \begin{equation}
  \langle i || V || j,nL \rangle  = \langle i || Y_L
  || j\rangle   \int dr V_0^T (r) u_i(r) u_j(r) \delta
  \rho^T_{nL},
  \end{equation}
which is exactly the form adopted in Ref.~\cite{Colo1994}.

Finally, we mention that, exactly in the same way as in
Ref.~\cite{Colo1994}, we introduce an isospin correction in the
matrix elements of $W^\downarrow$ in the $T_-$ channel. In fact, it
must be noted that in this case the coupling with the intermediate
states $|N\rangle$ can mix states with different isospins. These
intermediate states, which contain a phonon plus a proton particle
and a neutron hole, do not have pure isospin. They can be written as
 \begin{equation}
  | N \rangle = c_{-1} | N; T_0-1, T_0-1 \rangle + c_0 | N; T_0,
  T_0-1\rangle + c_{+1} |N; T_0+1, T_0-1 \rangle,
 \end{equation}
  where $T_0=(N-Z)/2$, and the coefficients $c_i$ are simply Clebsch-Gordan
  coefficients,
  \begin{eqnarray}
   c_{-1} &=& \langle 1 -1 T_0 T_0 | T_0-1, T_0 -1 \rangle =
   (2T_0-1)^{1/2} /(2T_0+1)^{1/2} , \nonumber\\
   c_0 &=& \langle 1 -1 T_0 T_0 | T_0, T_0 -1 \rangle =
   -(T_0+1)^{-1/2} , \nonumber\\
   c_{+1} &=& \langle 1 -1 T_0 T_0 | T_0+1, T_0 -1 \rangle =
   1/[(T_0+1)^{1/2} (2T_0+1)^{1/2}].
  \end{eqnarray}
(this writing is justified by the fact that our phonons,
as mentioned below, are to a good approximation isoscalar phonons).
The GT$^-$ resonance has isospin quantum numbers $|T,T_z\rangle = | T_0-1,T_0-1\rangle
 $, and its coupling with states of different isospin should be forbidden
 by the nuclear Hamiltonian. We impose that it is strictly forbidden
 (this amounts to neglecting Coulomb effects in the residual
 interaction) and, therefore, we project out the isospin
 component with the same value of the GT$^-$ resonance in the
intermediate states, i.e., the component $T=T_0-1$.
 Correspondingly, in the $W^\downarrow$ matrix elements, an isospin
 correction factor $|c_{-1}|^2 = (2T_0-1)/(2T_0+1)$  is added.

\section{Numerical details}\label{num}

In the present work, the HF equations are solved in coordinate space
on a radial mesh whose size is $0.1$ fm, within a box of $21$ fm.
The p-h configuration space used for the GT RPA calculation includes
all hole states and particle states (discretized in the mentioned
box) up to an upper cutoff $E_{\rm cut} = 100$ MeV. The results are
fully converged in this way (for example, the Ikeda sum rule of
$^{60}$Ni with the interaction SGII~\cite{Giai1981} reaches $11.998$
at the RPA level with this cutoff).

To build the model space of the p-h pairs plus phonon doorway
states, needed for $W^\downarrow$, the energies and reduced
transition probabilities of the most collective phonon modes with
spin and parity $2^+, 3^-, 4^+$ are calculated with the same energy
cutoff. With this choice in, e.g., $^{60}$Ni, the energy-weighted
sum rules (EWSRs) satisfy the double commutator values by about
$98\%$. Non-charge-exchange RPA is implemented in exactly the same
way as charge-exchange RPA, that is, with the same numerical input.

To minimize violations of the Pauli principle, and be consistent
with the very idea of particle-vibration coupling, only phonons
which absorb a fraction of the total isoscalar or isovector strength
larger than $5\%$ (and with energy less than $20$ MeV) are included
in the model space $Q_2$. The properties of the low-lying $2_1^+$,
$3_1^-$, and $4_1^+$ states in $^{60}$Ni calculated with three
different Skyrme interactions, namely SGII, SLy5, and SkM*, are
shown in Table~\ref{table1}.

%----------------------------------------------------------------------
\begin{table*}
\caption{Properties of low-lying phonons in $^{60}$Ni. The
theoretical results are calculated with the Skyrme interactions
SGII~\cite{Giai1981}, SLy5~\cite{Chabanat1998} , and
SkM*~\cite{Bartel1982}. The experimental data are taken from
Ref.~\cite{nndc}.}
\begin{tabular}{ccccccccc}
  \hline
  \hline
  State & \multicolumn{6}{c} {Theory}   & \multicolumn{2}{c} {Experiment} \\
            & \multicolumn{3}{c} {Energy} & \multicolumn{3}{c} {$B(EL, 0 \rightarrow L$ )} &
            Energy & $B(EL, 0 \rightarrow L)$ \\
            &\multicolumn{3}{c} { [MeV] } & \multicolumn{3}{c} {[e$^2$ fm$^{2L}$]} & [MeV]  & [e$^2$
            fm$^{2L}$] \\
           % \hline
            & SGII & SLy5 & SkM* &SGII & SLy5 & SkM* & &  \\
            \hline
 $2_1^+$      & $1.820$ & 1.157 & 2.212 &  $3.611\times 10^2$ &$3.562\times 10^2$ & $2.457\times 10^2$ &$ 1.333$  & $8.780
 \times 10^2$ \\
 $3_1^- $     & $4.897$ & 5.727 & 4.901 &$1.576 \times 10^4$ & $ 1.643 \times 10^4$&$1.150 \times 10^4$ & $4.040$ &  \\
 $4_1^+$    & $1.902 $ & 2.343 &3.469 & $ 1.007 \times 10^5$ &$ 7.257 \times 10^4$ &$ 6.365 \times 10^4$ &  $2.506$ &
   \\
 \hline
 \hline
 \end{tabular}
 \label{table1}
\end{table*}

%-------------------------------------------------------------------------------

We finally point out that we have checked, in the present
calculation of Gamow-Teller resonances, that the Ikeda sum rule is
still satisfied at the level of RPA plus particle-vibration
coupling. For instance, in the case of $^{60}$Ni calculated with the
Skyrme interactions SGII, SLy5, and SkM*, $m_0({\rm GT}^-) -
m_0({\rm GT}^+)=11.87,11.89$, and $ 11.83$, respectively, when the
strength is integrated up to $45$ MeV£»for $^{56}$Ni, the sum rule
value is -0.10 for the interaction SGII, -0.07 for SLy5, and -0.11
for SkM* when the strength is integrated up to 45 MeV; and for
$^{208}$Pb it is 129.29 for the interaction SGII when the strength
is integrated up to 70 MeV.

%------------------------------------------------------------------------------------------------
\section{Sensitivity of the RPA results to the choice of the Skyrme set}
\label{Skyrme}

We have first tried to understand the sensitivity of the
GT energies to the choice of the Skyrme interaction used
for RPA, and the underlying reasons.

The main components of the GT response are the transitions between
spin-orbit partners. In a nucleus such as $^{60}$Ni, which is our
benchmark in this study, the GT$^-$ (GT$^+$) response is expected to
be dominated by the single transition $\nu f_{7/2} \rightarrow \pi
f_{5/2}$ ($\pi f_{7/2} \rightarrow \nu f_{5/2}$). In such a case, we
can approximately write the energy of the GTR as
\begin{equation}\label{en_RPA}
E_{\rm RPA} = E_{\rm unper} + E_{\rm res} \approx \Delta E_{ls} + V.
\end{equation}
The first term $E_{\rm unper}$ is the unperturbed p-h energy, and
can be associated with the spin-orbit splitting $\Delta E_{ls}$
whereas the second term $E_{\rm res}$ is the energy shift induced by
the residual interaction and can be written in terms of its matrix
element $V$. The matrix elements of the residual interaction in the
spin-isospin channel are as a rule repulsive, so the two terms of
Eq.~(\ref{en_RPA}) have the same sign.

In the case of a Skyrme calculation, the first term is controlled by
the spin-orbit parameter $W_0$ (we leave aside the nonstandard
Skyrme sets that have two parameters in the spin-orbit part of the
energy functional, and those which include the so-called
$\mathbf{J}^2$ terms). The second term depends instead on the
strength of the residual interaction in the spin-isospin channel. In
general one imagines that it is mainly associated with the
well-known Landau parameter $g_0^\prime$, but we should not forget
that $g_1^\prime$ plays a role as well (these are the only two
Landau parameters that do not vanish for a zero-range interaction).
In Ref.~\cite{Fracasso2007}, by studying the strength of the GT$^-$
resonance in several nuclei including Sn isotopes and $^{208}$Pb, we
have concluded that this strength is sensitive to $g_0^\prime$ and
$g_1^\prime$, and that values around $0.45$ and $0.5$ are preferable
for these parameters.

%------------------------------------------------------------------------------------------------

\begin{table*}
\caption{Correspondence between numbers and interactions in
Fig.~\ref{fig2}, Fig.~\ref{fig3}, and Fig.~\ref{fig4}. }
\begin{tabular}{cccccccccccccc}
\hline \hline
Number &  1 & 2 & 3 & 4 & 5 & 6 & 7 & 8 & 9 & 10 & 11 & 12 & 13\\
 \hline
Interaction & SLy5 & SLy4 & SkP & LNS & SkI3 & SkM* & Sk255 & SIII &
BSk17 & SGII
 & SkO' & SkI4 & SkO\\
 Reference &  \cite{Chabanat1998} & \cite{Chabanat1998} &
 \cite{Dobaczewski1984} & \cite{Cao2006} & \cite{Reinhard1995} &
 \cite{Bartel1982} &  \cite{Agrawal2003} & \cite{Beiner1975} &
 \cite{Goriely2009} &  \cite{Giai1981} &  \cite{Reinhard1999} &
 \cite{Reinhard1995} & \cite{Reinhard1999} \\
  \hline \hline
\end{tabular}
 \label{table2}
\end{table*}

%---------------------------------------------------------------------------------------------------------
\begin{figure*}
\centerline{
\includegraphics[scale=0.35,angle=0]{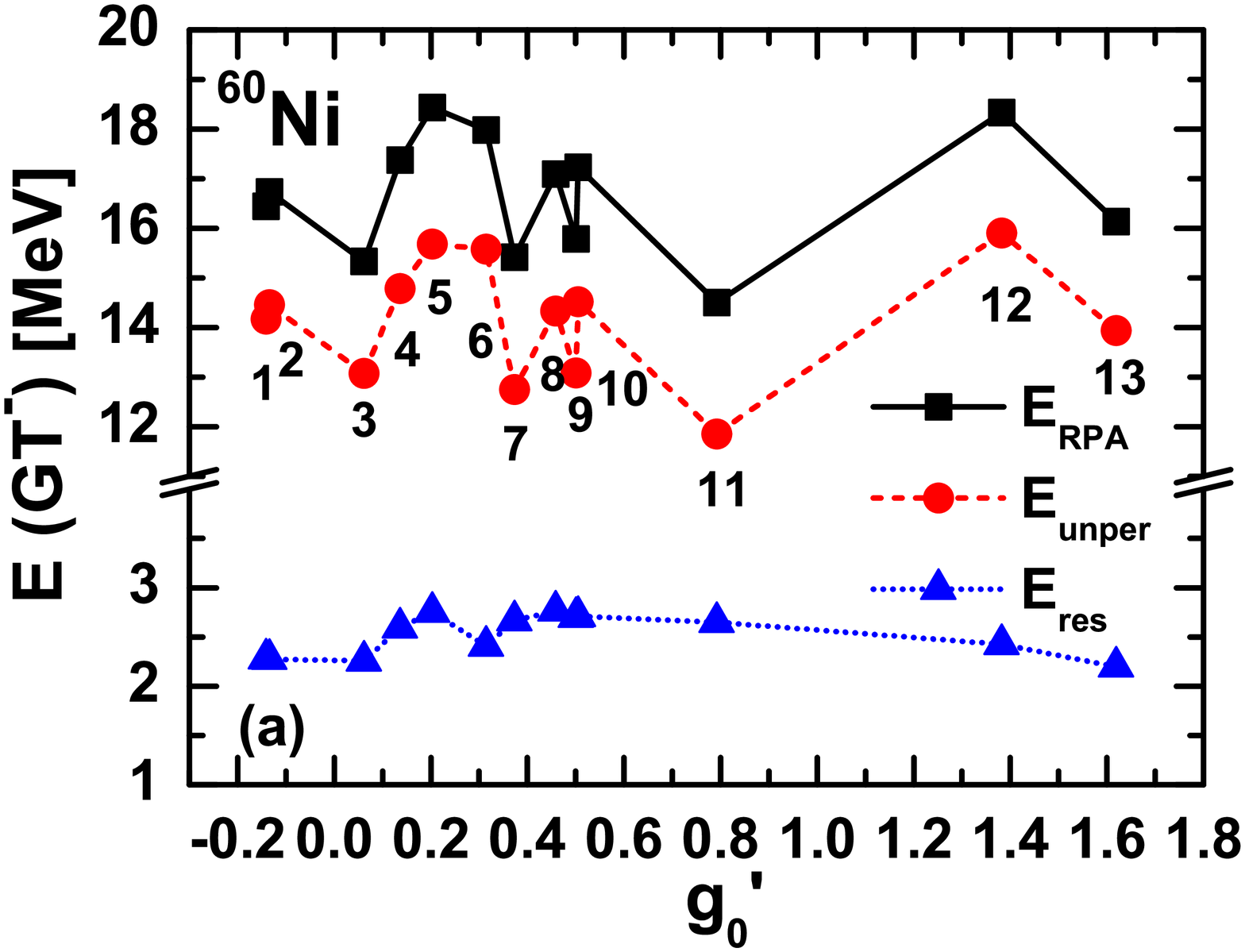}
\includegraphics[scale=0.35,angle=0]{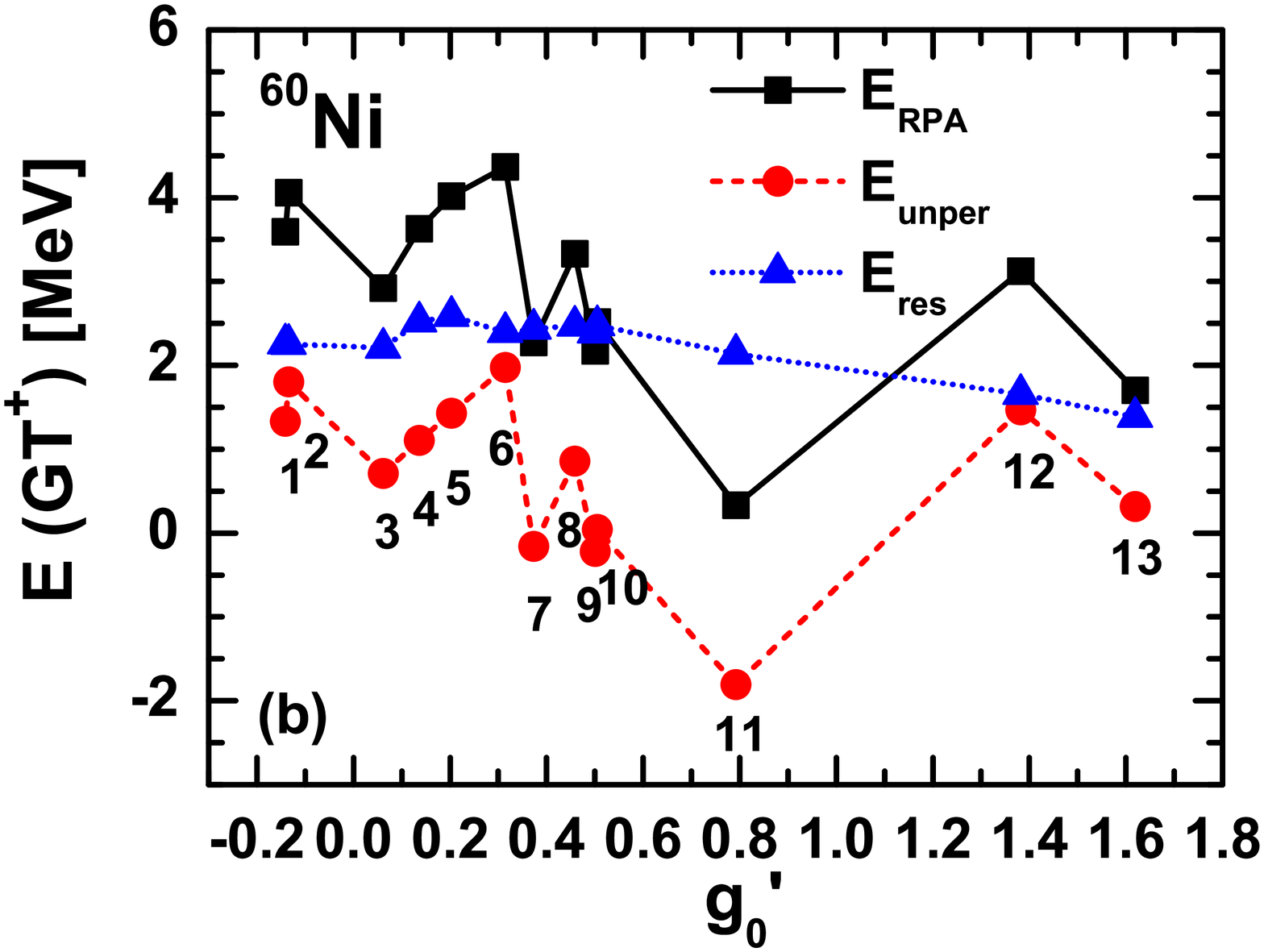}
} \caption{(Color online) The Gamow-Teller energy calculated in the
RPA ($E_{\rm RPA}$), the unperturbed energy of the main p-h
components ($E_{\rm unper}$, see the text), and the associated shift
$E_{\rm res} \equiv E_{\rm RPA} - E_{\rm unper}$ are displayed as
functions of the Landau parameter $g_0'$ for nucleus $^{60}$Ni. The
GT$^-$ and GT$^+$ cases are in panel (a) and panel (b),
respectively, and the correspondence between numbers and Skyrme sets
can be found in Table~\ref{table2}. } \label{fig2}
\end{figure*}
%----------------------------------------------------------------------------------------------------------

In the present study, we focus first on the energy position of both
the GT$^-$ and GT$^+$ peaks. In Fig.~\ref{fig2}, the following
quantities are displayed, as a function of $g_0^\prime$: the main
peak of the Gamow-Teller resonance calculated in RPA, the
unperturbed Gamow-Teller energy of the single p-h configuration
$(\pi f_{5/2})(\nu f_{7/2})^{-1} $ for GT$^-$ or $(\nu f_{5/2})(\pi
f_{7/2})^{-1} $ for GT$^+$, and the shift $E_{\rm res} \equiv E_{\rm
RPA} - E_{\rm unper}$. Surprisingly, the energy shift is flat as a
function of $g_0^\prime$: we have found that this is due to a strong
cancellation between the contributions associated with $g_0^\prime$
and $g_1^\prime$. The staggering of the GT$^-$ main peak is
associated with the staggering in the unperturbed energy (which is
obviously uncorrelated with the Landau parameters), and it follows
it quite closely. The same can be said of the GT$^+$ peak; in this
case the absolute value of the unperturbed energy is, however,
smaller than in the previous case and smaller than the typical
matrix element of the residual force.

%---------------------------------------------------------------------------------------------------------
\begin{figure*}
\centerline{
\includegraphics[scale=0.35,angle=0]{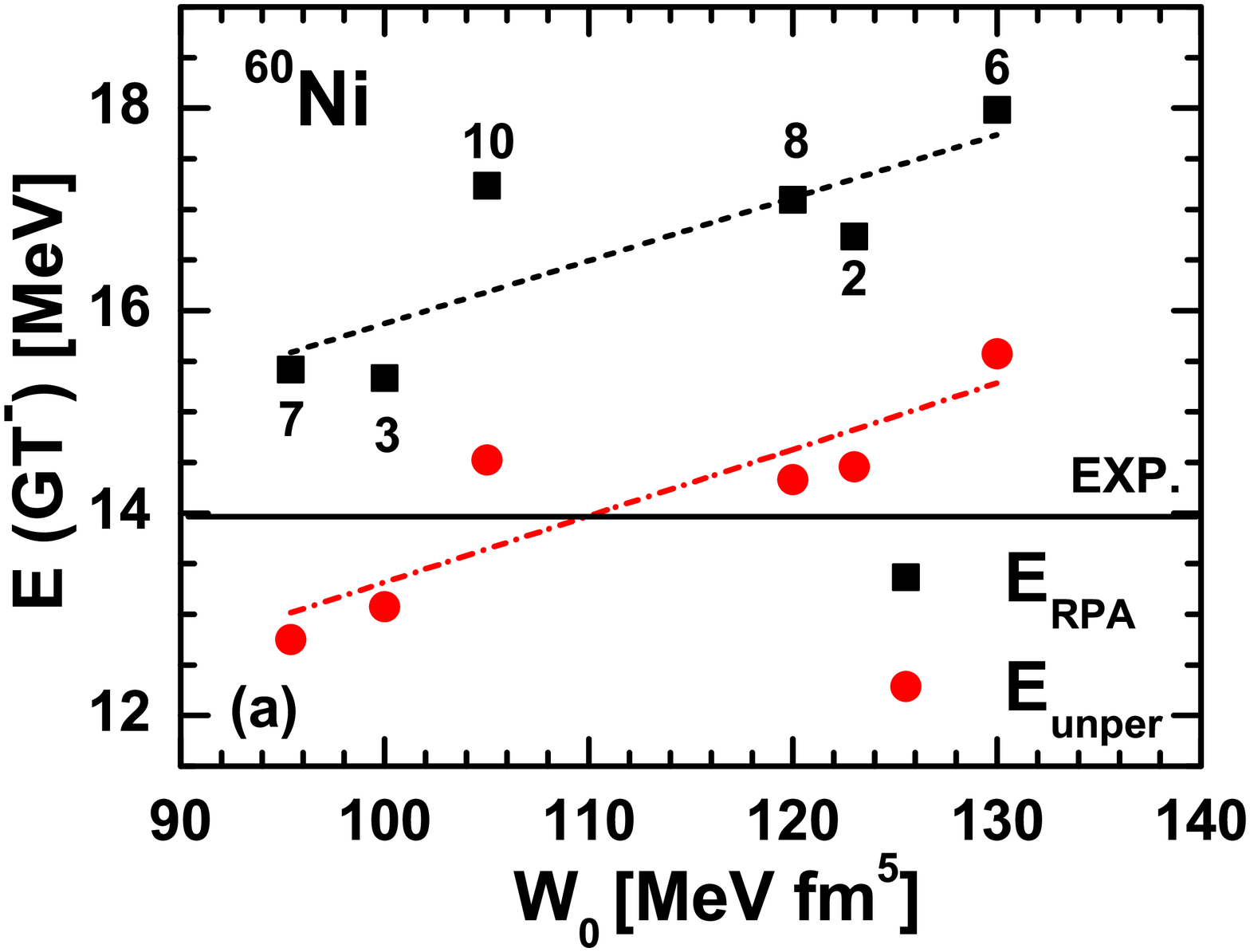}
\includegraphics[scale=0.35,angle=0]{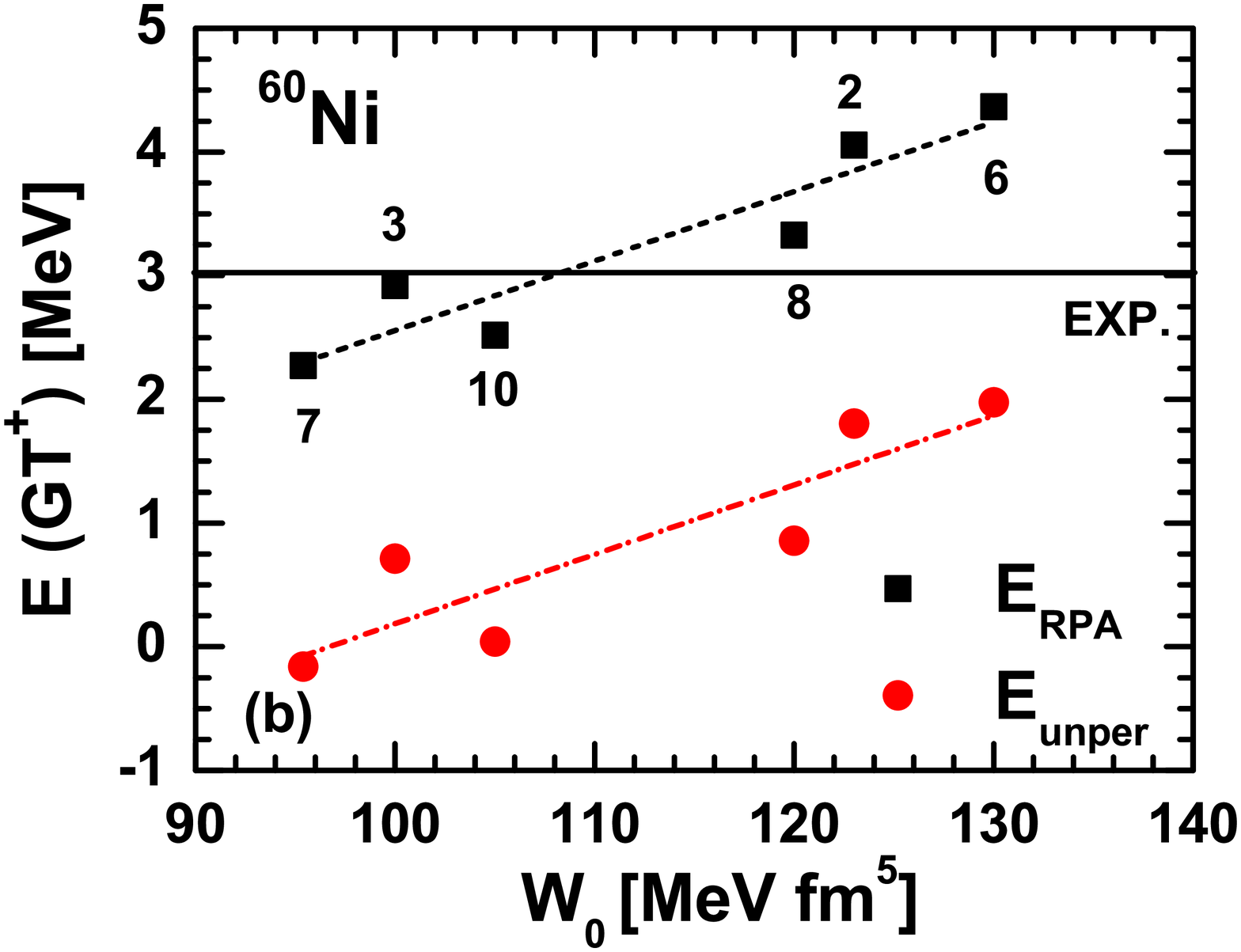}
} \caption{(Color online) The same energies $E_{\rm RPA}$ and
$E_{\rm unper}$ shown in Fig. \ref{fig2} are here displayed as
functions of $W_0$. The lines are linear fits that are, however,
shown only for illustrative purposes. See the text for a detailed
discussion. The experimental values for the energies (expressed with
respect to the parent nucleus as in the theory) are indicated by
means of black solid lines and are taken from
Ref.~\cite{Rapaport1983} for the GT$^-$ case and from
Ref.~\cite{Williams1995} for the GT$^+$ case. } \label{fig3}
\end{figure*}
%----------------------------------------------------------------------------------------------------------

In Fig.~\ref{fig3} the RPA and unperturbed energies are again
displayed as a function of the spin-orbit strength parameter $W_0$.
Only a subset of all the studied interaction has been chosen for
this figure. The dashed and dash-dotted lines are linear fits of the
points that are displayed only as guides to the eye. A {\em
quantitative} correlation between the parameter $W_0$ and the GT
energy can be neither inferred from the numerical results nor
expected, since the unperturbed energy is associated with a
neutron-proton transition. However, it is clear that there is a {\em
qualitative} tendency of both unperturbed and RPA energies to grow
as a function of $W_0$, and that these energies are more sensitive
to the single-particle spectrum than to other parameters such as
$g_0^\prime$ or $g_1^\prime$.

In the same figure we also display, by means of horizontal lines,
the experimental values of the energies taken from
Refs.~\cite{Rapaport1983,Williams1995}. Since the particle-vibration
coupling produces, as a rule, a downward shift of the resonance
energies (see the next section), we are inclined to prefer, for
further considerations, the sets marked as 2, 6, 8, and (at least
looking at the GT$^-$ case) 10. Set 8 is SIII, which is a force that
does not have very satisfactory properties as a whole (for instance,
the associated value of the nuclear incompressibility is too large).
Set 2 is SLy4, but SLy5 provides quite similar results. Sets 6 and
10 are, respectively, SkM* and SGII. We will discuss in the next
section some results obtained by using SkM*, SLy5, and SGII. SkM*
and SGII have reasonable values for the Landau parameters
\cite{Fracasso2007}.

%---------------------------------------------------------------------------------------------------------
\begin{figure*}
\centerline{
\includegraphics[scale=0.35,angle=0]{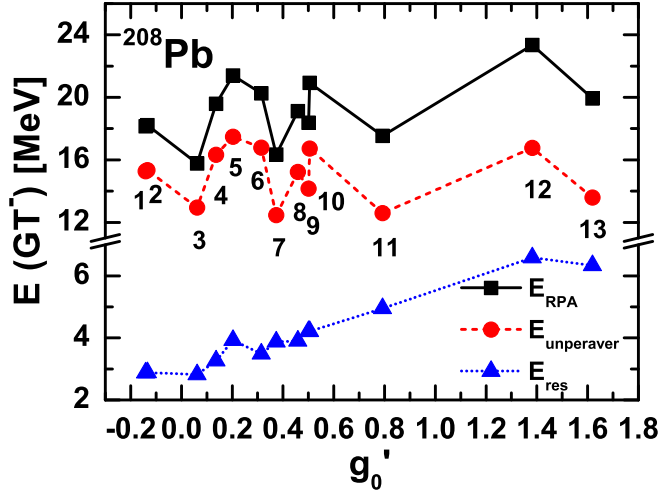}
} \caption{(Color online) The same quantities already displayed in
Fig. \ref{fig2} are now displayed in the case of $^{208}$Pb. Here,
the unperturbed energy $E_{\rm unperaver}$ is the weighted average
of the two main configurations $\nu i_{13/2} \rightarrow \pi
i_{11/2}$ and $\nu h_{11/2} \rightarrow \pi h_{9/2}$. } \label{fig4}
\end{figure*}
%----------------------------------------------------------------------------------------------------------

In Ref. \cite{Fracasso2007} it has been found, however, that the
parameters $g_0^\prime$ and $g_1^\prime$ are not irrelevant for the
GT properties, in keeping with the fact that they are correlated
with the GT$^-$ strength in $^{208}$Pb as well as in Sn isotopes. We
have checked here what happens for the GT$^-$ energy in $^{208}$Pb,
and our results are shown in Fig. \ref{fig4}. Although the
single-particle properties still play a relevant role -- as can be
seen from the staggering that is similar to that in the figure for
$^{60}$Ni -- here there is a clear tendency of the shift associated
with the residual interaction: it is not flat, rather it grows to
some extent with $g_0^\prime$. We have found that in fact there is
less cancellation between the terms associated with $g_0^\prime$ and
$g_1^\prime$. Moreover, in the schematic model, the RPA energy is
known to scale as
\begin{equation}\label{en_RPA_2}
E_{\rm RPA} \approx E_{\rm unperaver} + nV,
\end{equation}
where $n$ is the number of p-h configurations, $E_{\rm unperaver}$
is the weighted average of the unperturbed energies of the n
configurations contributing to the RPA state. This number is not 1
here, as it was in the case of $^{60}$Ni. The role of the residual
interaction must be larger here, accordingly.

In conclusion, the GT energy is, within the Skyrme framework, very
sensitive to the single-particle spectrum. This is especially true
in nuclei of the $fp$ shell (although we have discussed only
$^{60}$Ni in some detail, we have reached similar conclusions from
the analysis of $^{56,58}$Ni and $^{54,56}$Fe). There is some kind
of tendency for both the GT$^-$ and GT$^+$ energies to grow as a
function of the spin-orbit parameter $W_0$. Our analysis has shown
that a few Skyrme sets may be suitable to be employed for PVC
calculations with the expectation that the experimental energy can
be well reproduced.

\section{RPA with particle vibration coupling}
\label{PVC}

%---------------------------------------------------------------------------------------------------------
\begin{figure*}
\centerline{
\includegraphics[scale=0.35,angle=0]{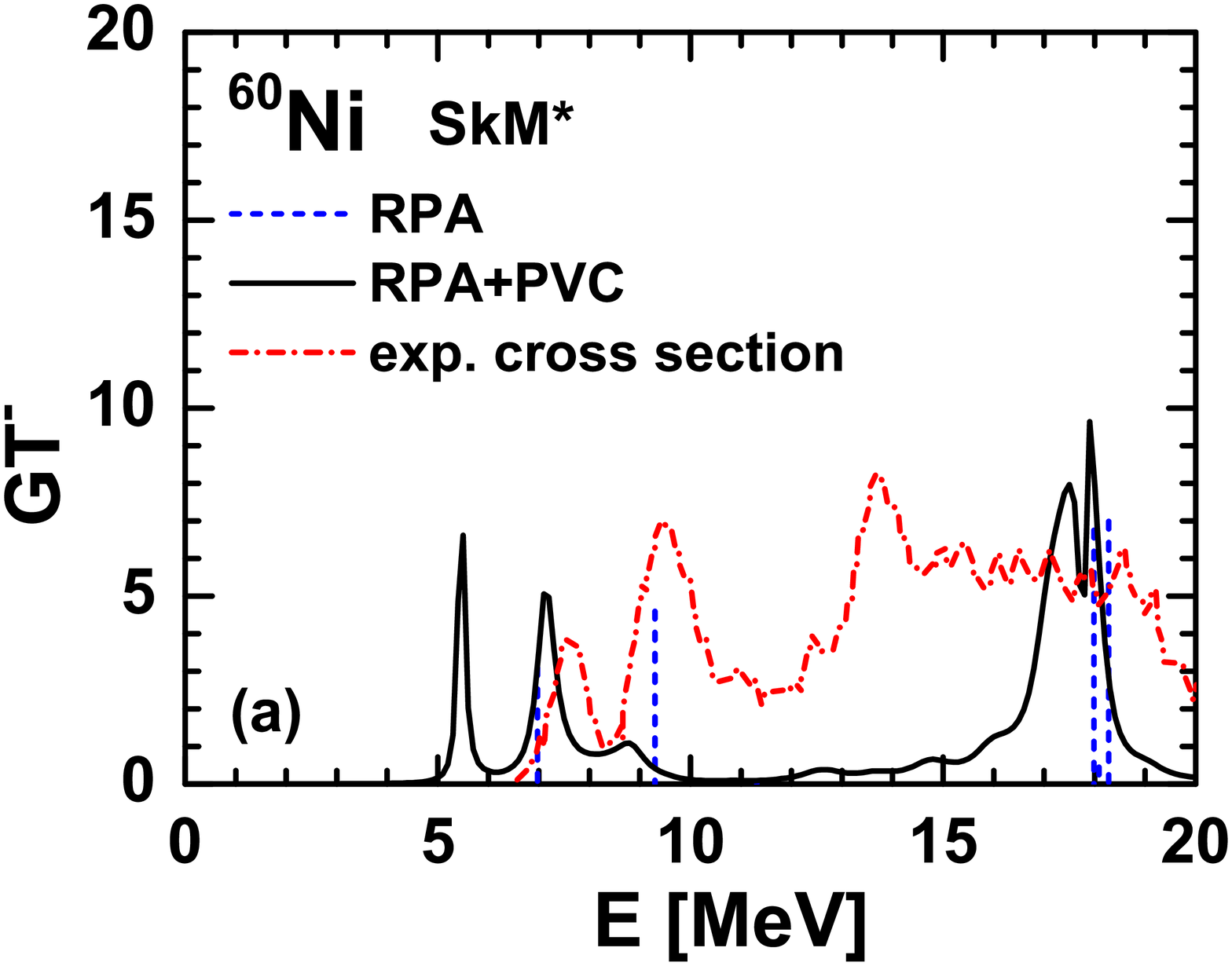}
\includegraphics[scale=0.35,angle=0]{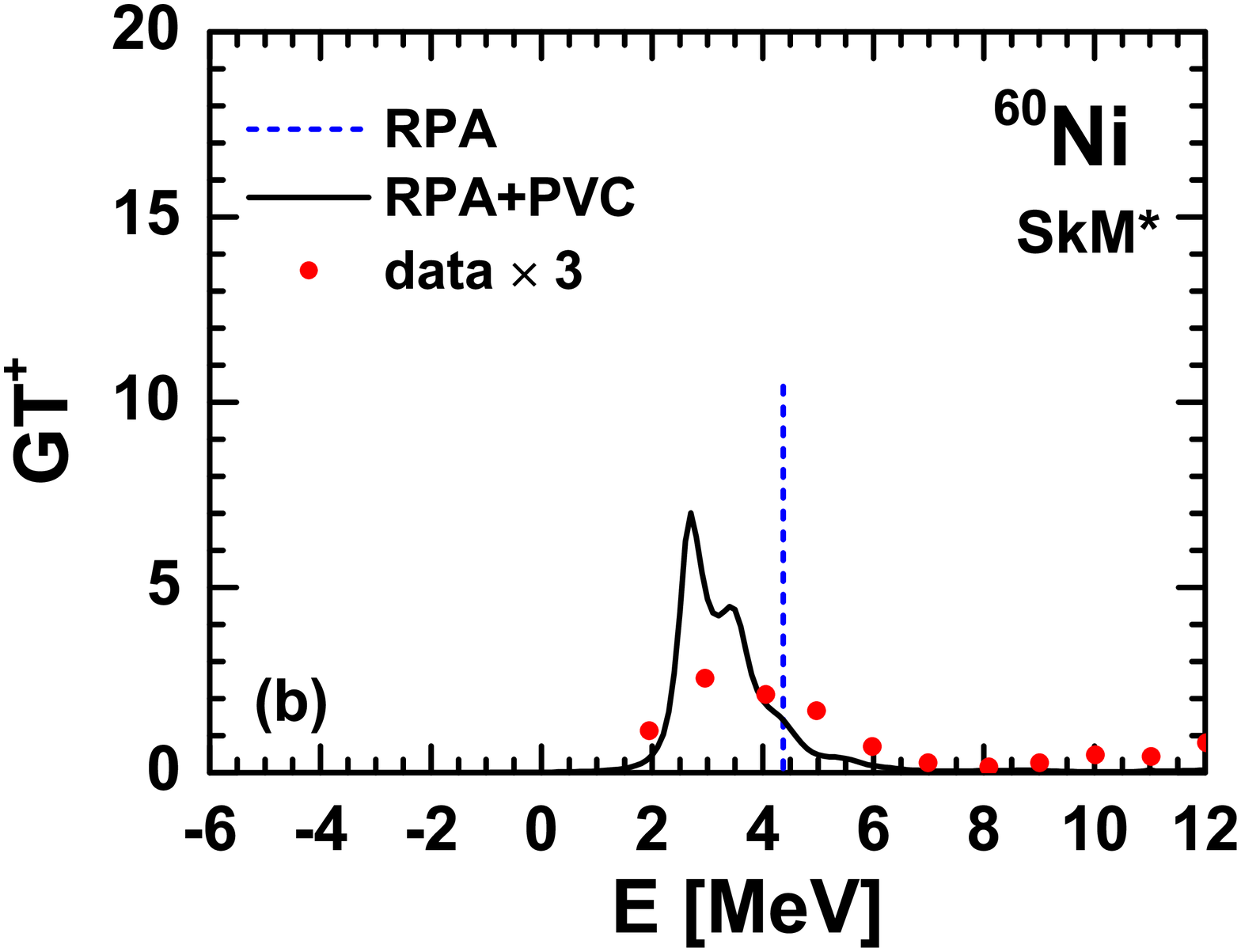}
} \caption{(Color online) Gamow-Teller strength distributions
calculated with the Skyrme interaction SkM* for the nucleus
$^{60}$Ni. The blue dashed discrete lines denote the RPA strength
(with dimensionless units), and the black solid lines represent the
distribution (with units of MeV$^{-1}$) calculated by the RPA+PVC
model. The experimental data for the case of
GT$^-$~\cite{Rapaport1983} [with units of mb/(sr~MeV)] and
GT$^+$~\cite{Williams1995} (with units of MeV$^{-1}$), displayed by
either a red dash-dotted line or red points, are discussed in the
text. } \label{fig5}
\end{figure*}
%----------------------------------------------------------------------------------------------------------

In the present section we discuss the results of the calculations
that include the coupling with vibrations. We use the interactions
SkM*, SLy5, and SGII, following the discussion in the previous
section. The goal is to see if, and to what extent, the coupling
with vibrations decreases the energies (in particular the GT$^+$
threshold energy) and changes the distribution of the strength. Of
course we are also interested in seeing if RPA plus PVC can
reproduce the total resonance width. In our previous work for
$^{208}$Pb \cite{Colo1994}, we have seen that the coupling with
vibrations can account for most of the spreading width of the GTR.

\subsection{Energies and widths of the GT states in $^{60}$Ni}

\begin{figure*}
\centerline{
\includegraphics[scale=0.35,angle=0]{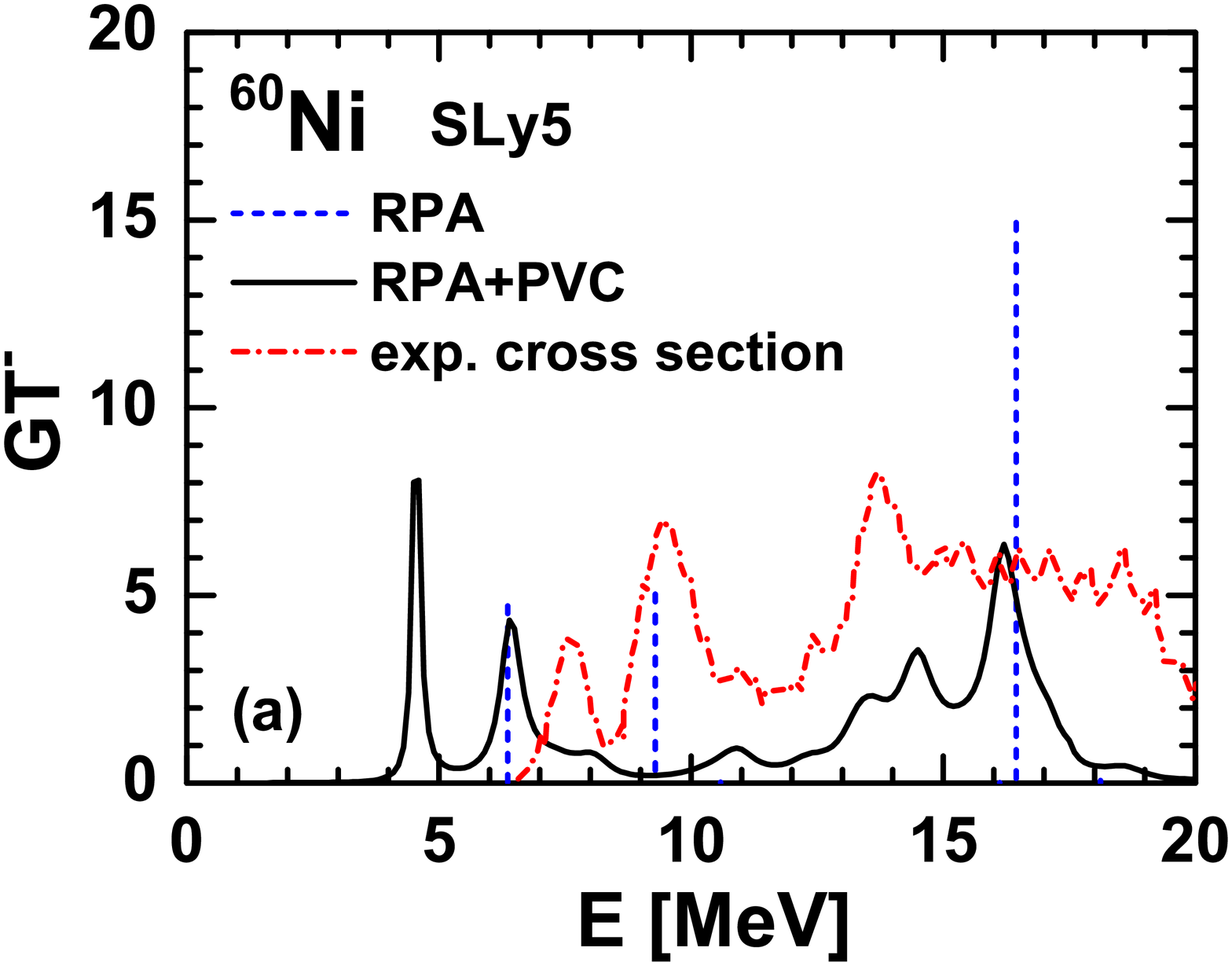}
\includegraphics[scale=0.35,angle=0]{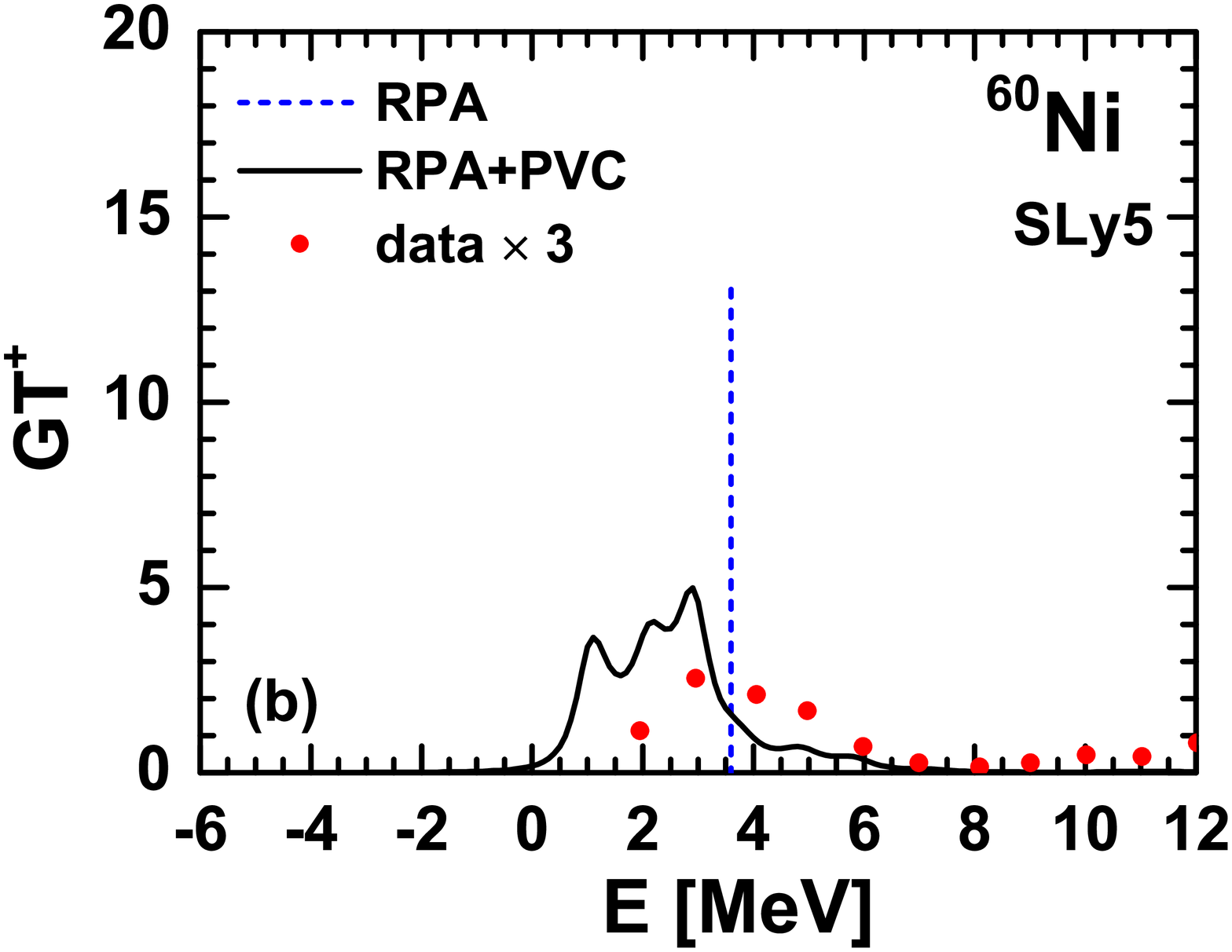}
} \caption{(Color online) The same as Fig.~\ref{fig5} in the case of
the Skyrme interaction SLy5.} \label{fig6}
\end{figure*}

The Gamow-Teller strength distributions as a function of the energy
(with respect to the parent nuclei) are displayed in
Figs.~\ref{fig5},~\ref{fig6}, and~\ref{fig7} in the case of the
Skyrme interactions SkM*, Sly5, and SGII, respectively, for the
nucleus $^{60}$Ni. The blue dashed discrete lines denote the RPA
strength, and the black solid lines represent the strength
distribution calculated by the RPA+PVC model. From all figures, it
is clear that the inclusion of phonon coupling leads to two
significant effects for both the GT$^-$ and the GT$^+$ case: there
is a downward shift of the excitation energy, and the development of
a spreading width. Especially in the high-energy region, a single
RPA GT$^-$ peak spreads into a wide resonance covering an energy
region of the order of 5 MeV (or more).

The development of a realistic spreading width allows, in principle,
a full comparison of the theoretical results with the experimental
data. The spreading width is caused by the coupling with the
phonons, through the imaginary part of $W^\downarrow$. In the case
of GT$^-$, a detailed strength distribution is not found in Ref.
\cite{Rapaport1983}, so only the measured zero-degree cross section
is displayed in Figs.~\ref{fig5},~\ref{fig6} and~\ref{fig7} by means
of a dash-dotted red line (cf. the left panels). All three
interactions produce two peaks in the low-energy part, and one main
high-lying GTR. These qualitative features are in agreement with the
experimental findings. However, the energies of the low-lying peaks
are underestimated, and the GTR is not as broad as in the
experiment. One can say we can account for a relevant fraction of
the measured total width, yet the high-lying experimental cross
section looks extremely flat.

In the case of the GT$^+$ the experimental data from
Ref.~\cite{Williams1995} are indicated by points. The total width of
the GT$^+$ distribution is about $3$ MeV. The theoretical result
obtained with the force SkM* agrees very well with experiment, as
far as both the peak energy and the width are concerned. SLy5 also
reproduces well the width, whereas in SGII one finds two peaks but
still a fragmentation of the strength over a reasonable energy
interval. It must be noted that the experimental data are multiplied
by a factor of 3 in the right panels of Figs.~\ref{fig5},~\ref{fig6}
and~\ref{fig7}. We discuss this issue in the next subsection.

\subsection{The quenching problem}

As a result of the phonon coupling, we can say we redistribute the
Gamow-Teller strength in a region that is several MeV wide. However,
this does not produce a real quenching. In the case of $^{60}$Ni
with the force SGII, up to the energy of $20$ MeV ($10$ MeV), the
integrated GT$^-$ (GT$^+$) strength calculated by RPA+PVC is $19.4$
($8.1$). For comparison, the RPA value is $20.8$ ($9.3$).
Consequently, the strength is quenched by $7\%$ ($13\%$) compared
with RPA but it is larger than the experimental data. In
experimental results, one finds a total strength equal to $7.2 \pm
1.8$ in the case of the GT$^-$~\cite{Rapaport1983} (up to $21$ MeV
in the parent nucleus), and $3.11 \pm 0.08$ in the case of the
GT$^+$~\cite{Williams1995} (up to $10$ MeV in the parent nucleus).
In the calculation of RPA+PVC with interaction SGII, only $15\%$ of
the sum rule shifts to energies above 20 MeV.

SkM* and SLy5 give similar (i.e., small) quenching effects as SGII.
In the case of SLy5, one finds an integrated strength of $23.4$ up
to $20$ MeV for the GT$^-$ case after coupling with phonons (instead
of $24.8$ given by RPA). For the GT$^+$ case, the integrated
strength up to $10$ MeV in RPA+PVC is $12.1$ (instead of $13.0$ from
RPA). $11\%$ of the total sum rule is shifted to the energy region
$20- 45$ MeV. In the case of SkM*, the integrated strength for
GT$^-$ (GT$^+$) from RPA+PVC is $20.3$ ($9.5$) up to $20$ MeV ($10$
MeV), instead of $21.9$ ($10.4$) from RPA. $15\%$ percent of the sum
rule is shifted to the energy region $20-45$ MeV.

There are probably two missing effects that can explain the lack of
quenching in our calculations (let alone the small, or negligible,
coupling with the $\Delta$ isobar). In Ref. \cite{Bertsch1982} a
large shift of strength to high energies is found by coupling with
uncorrelated 2p-2h doorway states. These extend up to higher energy,
as compared with the 1p-1h plus phonon doorway states. Consequently,
although less important for the spreading width, they can account
for the shift of strength above 20 MeV. In that work, the tensor
force has been found to play a role as well. Even in the limited
1p-1h space, the tensor force has been found to be capable of
shifting strength at high energy, although in smaller amounts
\cite{Bai2009}.

%------------------------------------------------------------------

\subsection{Microscopic origin of the downward
shift induced by phonon coupling}

\begin{figure*}
\centerline{
\includegraphics[scale=0.35,angle=0]{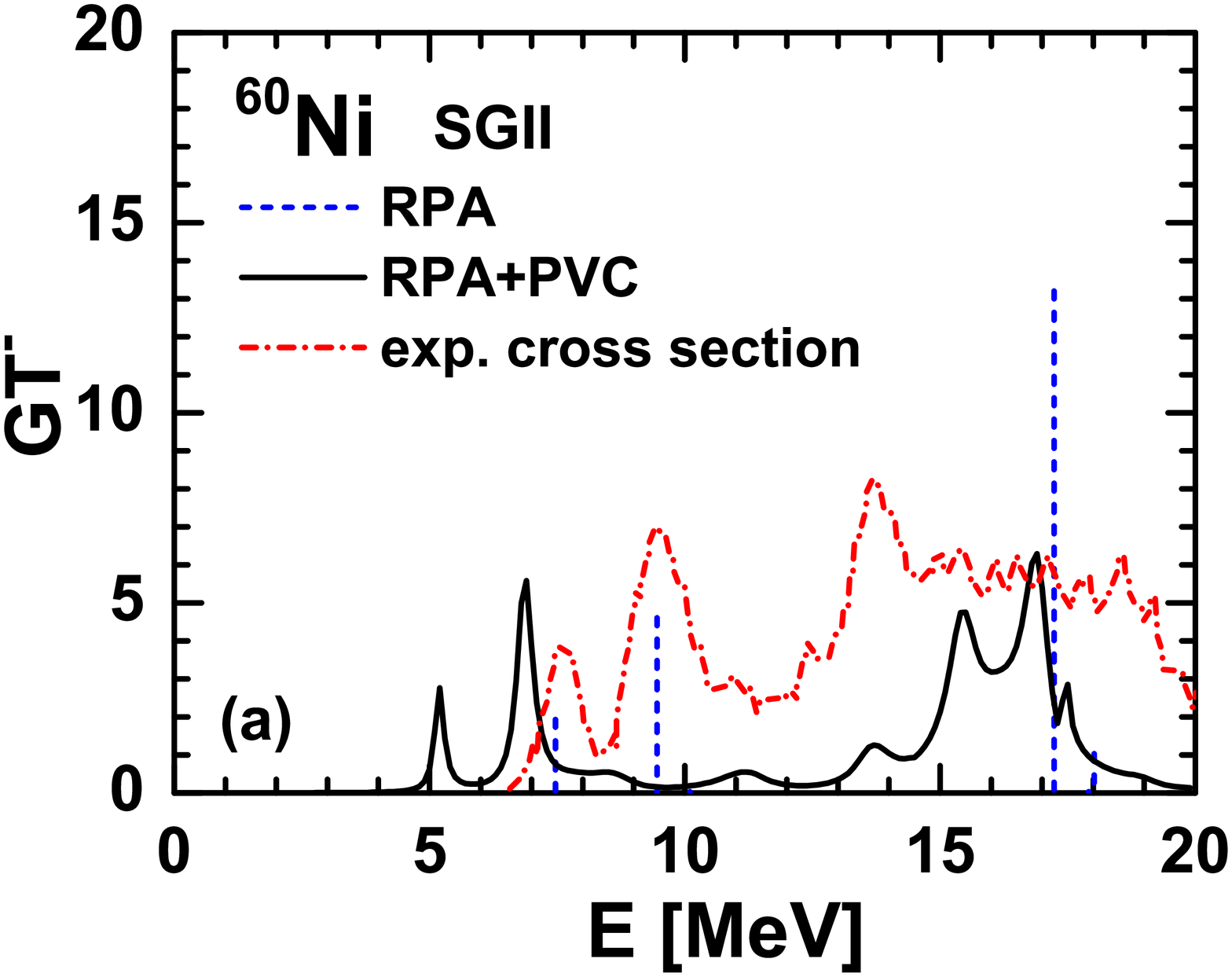}
\includegraphics[scale=0.35,angle=0]{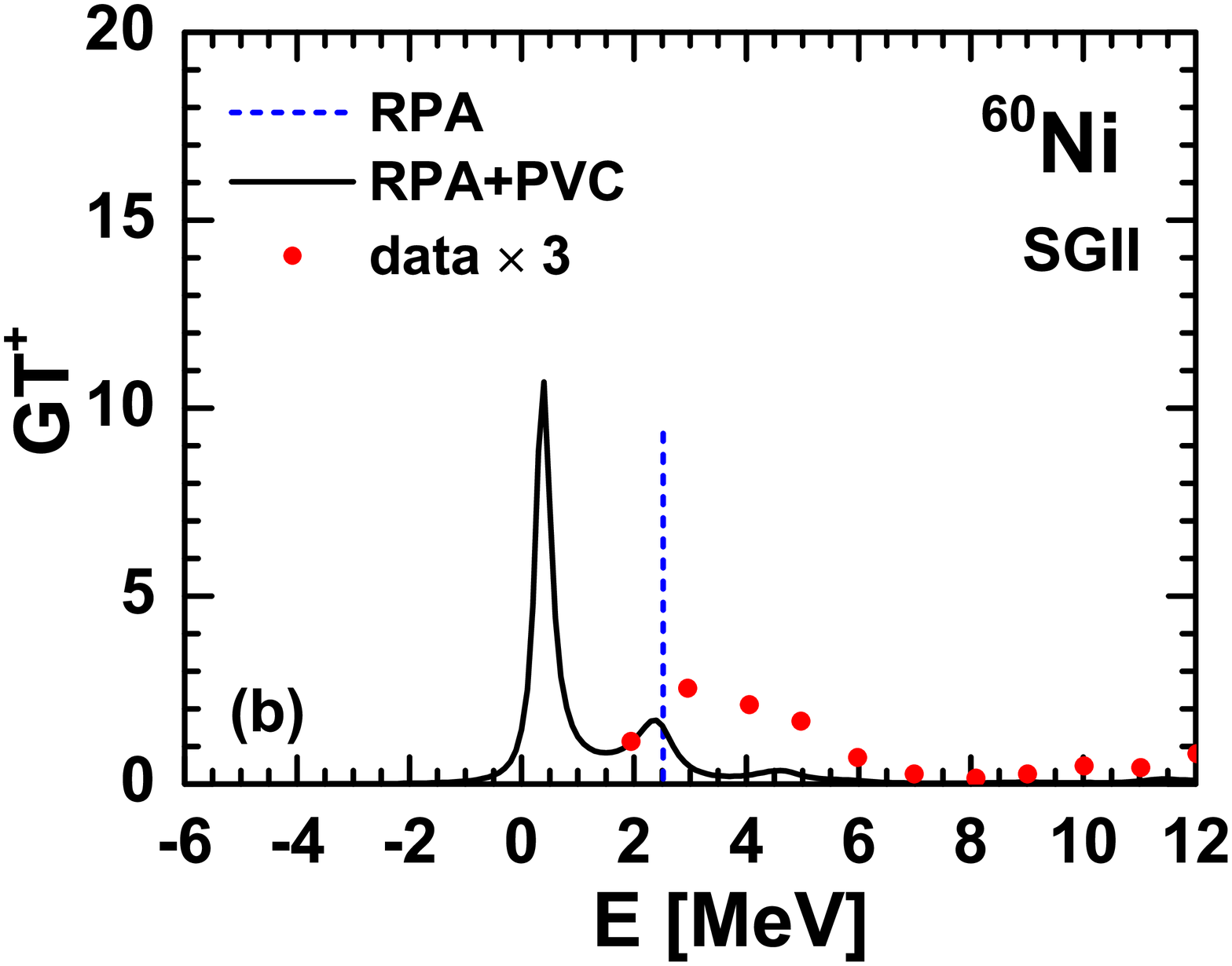}
} \caption{(Color online) The same as Fig. \ref{fig5} in the case of
the Skyrme interaction SGII. } \label{fig7}
\end{figure*}

In general, it can be expected that the coupling with phonons leads
to a downward shift of the resonance peaks, and that the low-lying
vibrations are the most effective. We analyze here in detail the
effects in the case of the GT$^+$ peak of Fig. \ref{fig7}.

The total shift in the peak energy for the GT$^+$ channel is $-2.1$
MeV, and the $2^+, 3^-, 4^+$ phonons give partial contributions
equal to $-1.3, -0.8$, and $-0.1$ MeV, respectively. For every
multipole, the lowest phonon, which is the most collective one,
plays the most important role, giving $85\%$ ($-1.1$ MeV) and $75\%$
($-0.6$ MeV) of the shift for the $2^+$ and $3^-$ case,
respectively. The diagonal matrix element $W^\downarrow_{ph,ph}$,
when the $ph$ configuration is the dominant component of the GT$^+$
peak [$\nu 1f_{5/2} (\pi 1f_{7/2})^{-1}$], is identified as the
dominant term responsible for the energy shift. In the case of
coupling with the first $2^+$ phonon, the first [Fig.~\ref{fig1}
(1)] and second [Fig.~\ref{fig1} (2)] diagrams of
$W^\downarrow_{ph,ph}$ give similar negative contributions to the
energy shift both with values of about $-1.3$ MeV, while the third
[Fig.~\ref{fig1} (3)] and fourth [Fig.~\ref{fig1} (4)] diagrams
cancel them by about $70\%$. These numbers are calculated at the
value of $\omega$ at which the GT peak lies if we couple the RPA
state with the 2$^+_1$ state only, namely $\omega = 1.4$ MeV. In the
first and second diagrams of Fig.~\ref{fig1}, the $2^+$ phonons can
only couple to the intermediate states with negative parity. It
turns out that the particle states $\nu 1f_{5/2}$ and $\nu 2p_{1/2}$
are the most important intermediate states for the first graph, and
provide contributions accounting for $62\%$ ($-0.8$ MeV) and $23\%$
($-0.3$ MeV) of the total value of the first diagram ($-1.3$ MeV),
while the intermediate hole state $\pi 1f_{7/2}$ is the most
important one for the second graph, giving almost all the
contribution to the total value of the second diagram ($-1.3$ MeV).
In the case of coupling with the first $3^-$ phonons, the first and
second diagrams still give similar negative contributions to the
energy shift (both about $-0.3$ MeV, these values being calculated
at the value of $\omega$ at which the GT peak lies if we couple the
RPA state with the 3$^-_1$ state only, namely $\omega=1.9$ MeV);
however, the third and fourth diagrams are zero in the case of
$W^\downarrow_{ph,ph}$ when $ph$ is $\nu 1f_{5/2}(\pi
1f_{7/2})^{-1}$, due to parity conservation. In this matrix element,
the $3^-$ phonons only couple with intermediate states with positive
parity. The states $\nu 1g_{9/2},\nu 1g_{7/2}, \nu 2d_{5/2}$ are the
most important particle states and the associated contributions to
the first diagram of Fig.~\ref{fig1} are $-0.05$, $-0.05$, and
$-0.03\ {\rm MeV}$ (the total value of the first diagram is $-0.3$
MeV). Similarly, in the second diagram the important hole states are
$\pi 2s_{1/2}, \pi 1d_{5/2}, \pi 1d_{3/2}$, and the contributions
are $-0.13$, $-0.08$, and $-0.05\ {\rm MeV}$ (the total value of the
second diagram is $-0.3$ MeV).

%---------------------------------------------------------------------------------------------------------
%----------------------------------------------------------------------------------------------------------

%---------------------------------------------------------------------------------------------------------
%----------------------------------------------------------------------------------------------------------

%---------------------------------------------------------------------------------------------------------
\begin{figure*}
\centerline{
\includegraphics[scale=0.35,angle=0]{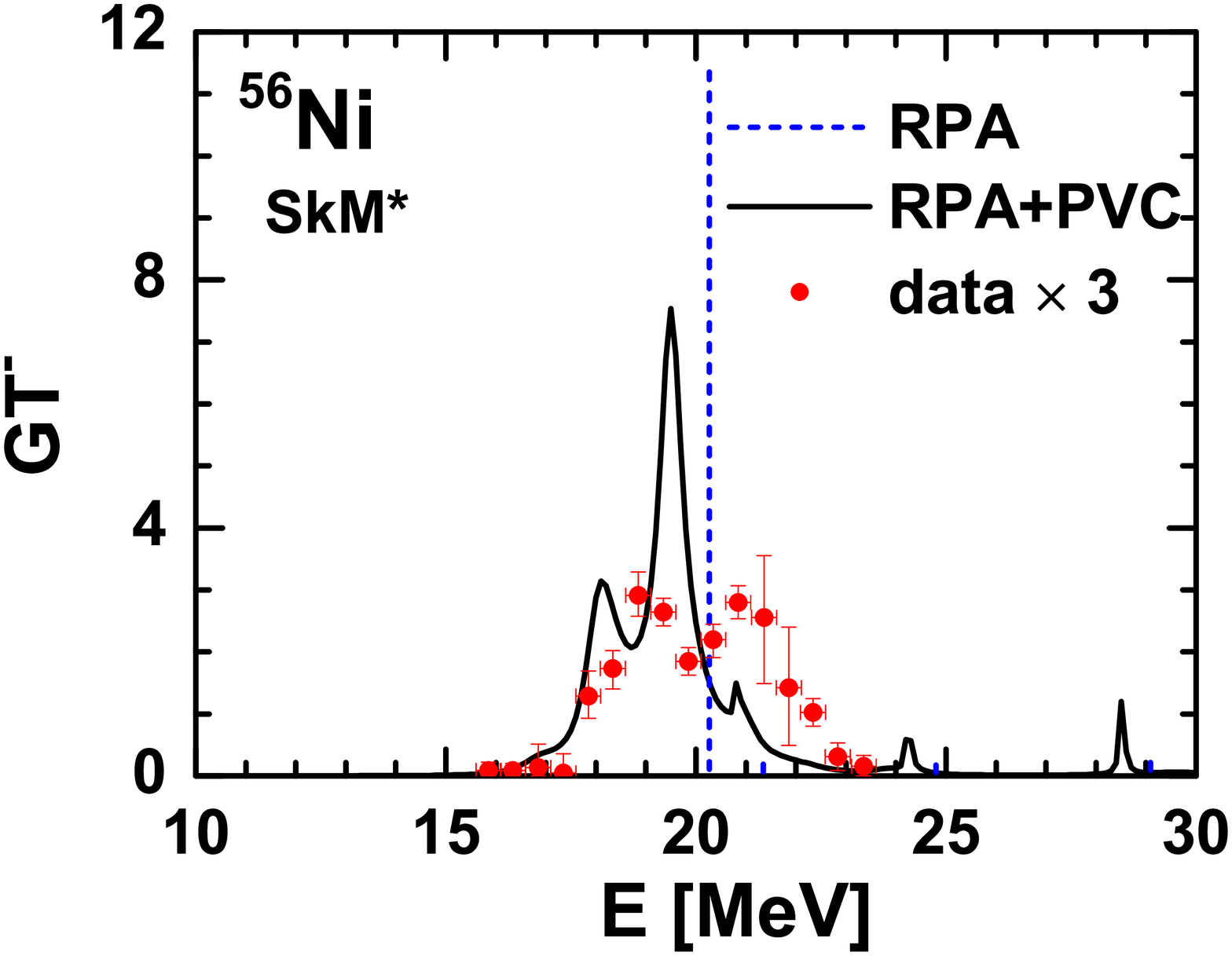}
} \caption{(Color online) Gamow-Teller strength distributions
calculated with the Skyrme interaction SkM* for the nucleus
$^{56}$Ni. The blue dashed discrete line denotes the RPA strength
(with dimensionless units), and the black solid line represents the
distribution (with units of MeV$^{-1}$) calculated by the RPA+PVC
model. The experimental data (with units of MeV$^{-1}$) from Ref.
\cite{Sasano2011} are indicated by red points. } \label{fig8}
\end{figure*}
%----------------------------------------------------------------------------------------------------------

%---------------------------------------------------------------------------------------------------------
\begin{figure*}
\centerline{
\includegraphics[scale=0.35,angle=0]{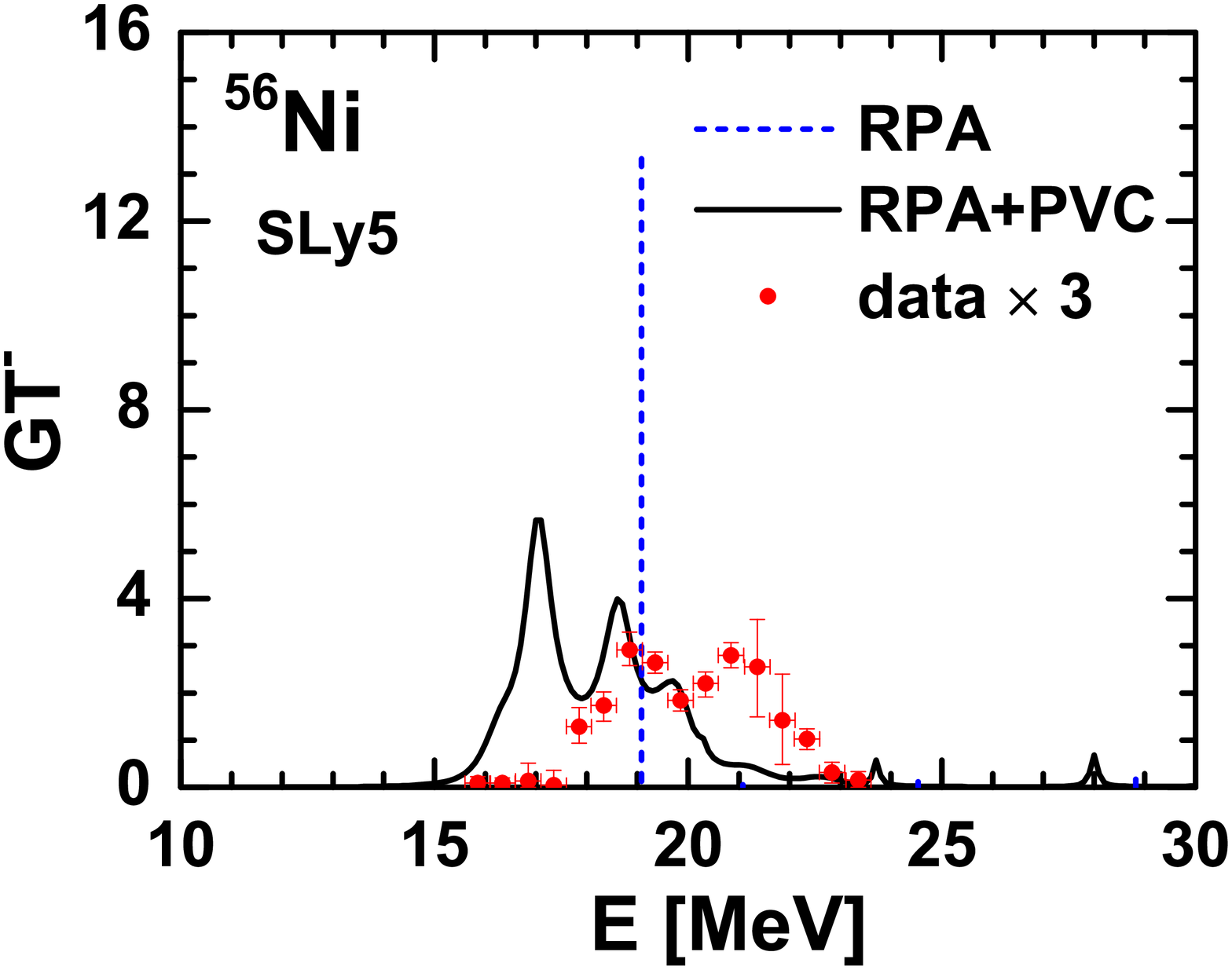}
} \caption{(Color online) The same as Fig. \ref{fig8} in the case of
the Skyrme interaction SLy5. } \label{fig9}
\end{figure*}
%----------------------------------------------------------------------------------------------------------

%---------------------------------------------------------------------------------------------------------
\begin{figure*}
\centerline{
\includegraphics[scale=0.35,angle=0]{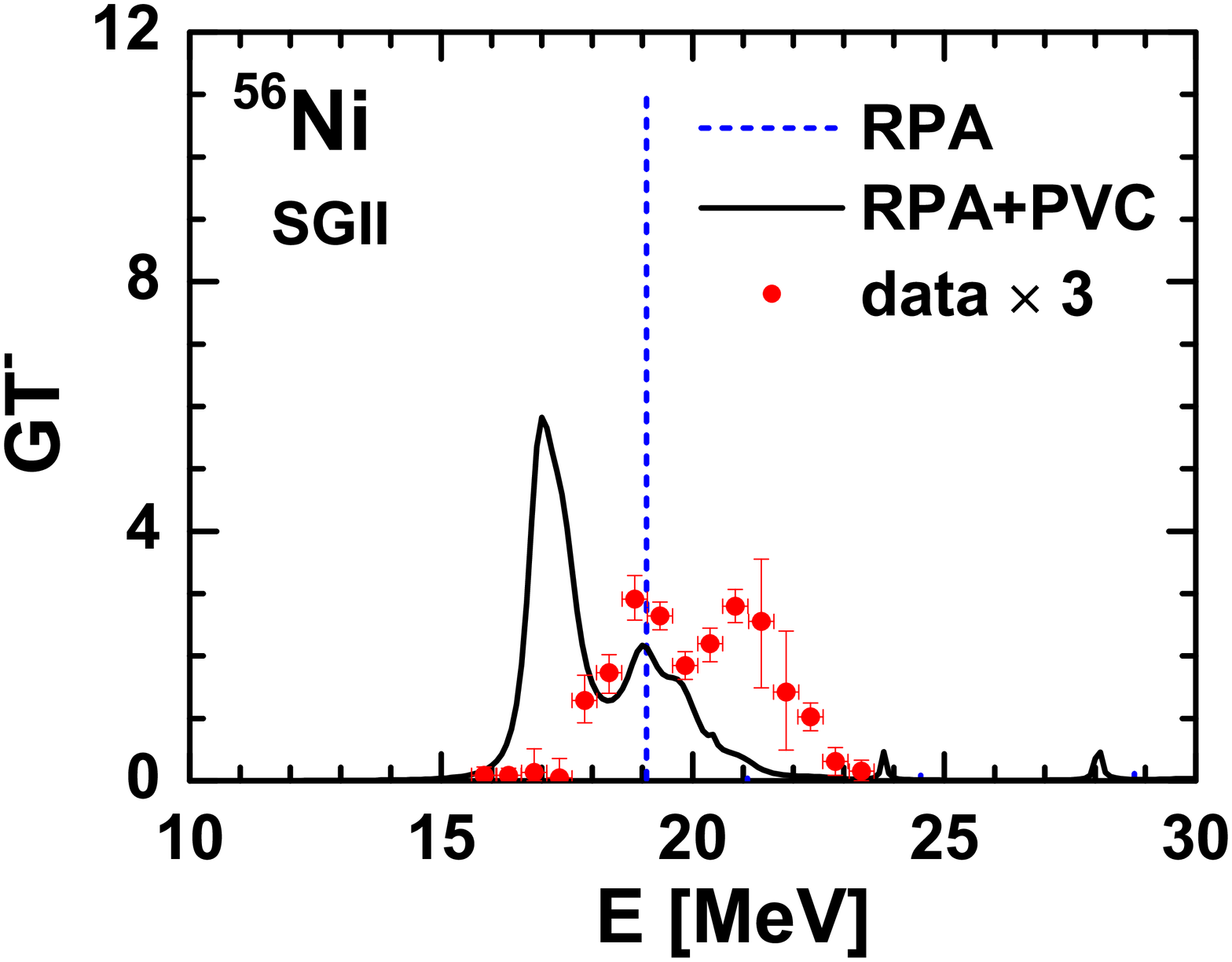}
} \caption{(Color online) The same as Fig. \ref{fig8} in the case of
the Skyrme interaction SGII. } \label{fig10}
\end{figure*}
%----------------------------------------------------------------------------------------------------------

\subsection{Results for the nucleus $^{56}$Ni}

Since the GT$^-$ resonance in the nucleus $^{56}$Ni was recently
measured~\cite{Sasano2011}, we also present results for this
nucleus, calculated with the three interactions SkM*, SLy5, and SGII
(cf. Figs.~\ref{fig8},\ref{fig9}, and \ref{fig10}, respectively).
Similarly to $^{60}$Ni, the coupling with phonons shifts the peaks
downward and produces a spreading width, yet brings a quite limited
quenching effect. The energy shift caused by the phonon coupling is
about $1-2$ MeV for all three interactions. The energy calculated by
means of SGII and SLy5 is somewhat lower than the experimental data,
while the energy from SkM* agrees very well with them. In this
latter case (SkM*) the strength spreads out in the same energy
region as the experimental strength distribution. In experimental
results, there is a double-peak structure and each peak has a width
of about $2$ MeV. In our calculations, we also obtain a similar
structure in the strength distribution for all three interactions.
For SGII and SLy5, the first peak has a width of about $1$ MeV and
the second peak has a width of about $2$ MeV. For SkM*, the width is
about $1.5$ MeV for the first peak and $1$ MeV for the second one.
The big difference between our results and the experimental data is
the absolute value of the strength. As in $^{60}$Ni, the coupling
with phonons does not induce a significant quenching (see the above
discussion). In the energy region $10-24$ MeV, after the coupling
with phonons, the integrated strength becomes $10.3, 12.8,$ and
$10.8$ (instead of $11.0, 13.4$, and $11.5$ calculated in RPA), for
the interactions SGII, SLy5, and SkM*, respectively. The same value
is $3.5 \pm 0.3 \pm 1.0$ from experiment~\cite{Sasano2011}.

%---------------------------------------------------------
\section{Conclusion}\label{conclu}
%---------------------------------------------------------

In the present work, we have analyzed the features of the
Gamow-Teller strength distributions in nuclei of the $fp$ shell.
These nuclei are of interest for astrophysical applications, but few
have been also objects of experimental measurements. Our purpose is
to improve existing models so that they can reproduce the
experimental findings, whenever available, and become reliable for
calculations of the electron capture rates needed for astrophysical
simulations; our ambition is to avoid resorting to free, {\em ad
hoc} parameters.

We have first performed RPA calculations with many different Skyrme
parameter sets. We have seen that the energy of the GT$^-$ and
GT$^+$ peak is quite sensitive to the single-particle properties and
in particular, to a large extent, to the spin-orbit strength
parameter. The energy shift caused by residual interaction is less
relevant to the Landau parameter $g_0'$ for these medium-mass nuclei
than in the case of, e.g., $^{208}$Pb, the reason being the smaller
number of p-h configurations.

Coupling with phonons is relevant to producing a more realistic
strength distribution, characterized by a spreading width. The
interaction SkM* does reproduce the peak position and the spreading
width, in the cases of the GT$^+$ resonance in $^{60}$Ni and the
GT$^-$ resonance in $^{56}$Ni, after the phonon coupling is taken
into account. This is important for the astrophysical applications
since the GT$^+$ energy is associated with the threshold energy for
electron capture. Other interactions display nonetheless, in the
same nuclei, realistic strength fragmentation after the phonon
coupling is taken into account.

We omit a full description of the GT strength quenching. While in
shell-model calculations this is often accounted for by means of a
free parameter, we refrain from making this choice here. Our model
probably can be improved by including coupling with doorway states
extending to higher energy, and by taking care of the tensor force
as well.

%=================================================================
{\center{\bf ACKNOWLEDGMENTS}}

This work was partly supported by the Major State 973 Program No.
2007CB815000, by the National Natural Science Foundation of China
under Grants No. 10975008 and No. 11175002, and by the Research Fund
for the Doctoral Program of Higher Education under Grant No.
20110001110087. The support of the Italian Research Project
``Many-body theory of nuclear systems and implications on the
physics of neutron stars" (PRIN 2008) is also acknowledged, as it
has allowed the visit of one of the authors (Y.N.) in Milano.

%=================================================================

%
%------------------------------------------------------------------

\clearpage
%\bibliographystyle{apsrev}

%\end{CJK*}
\end{document}